\documentclass[sigconf,nonacm]{acmart}
\usepackage[T1]{fontenc}
\usepackage{amsmath}
\usepackage{natbib}
\usepackage{graphicx}
\usepackage{booktabs}
\usepackage{multirow}
\usepackage{hyperref}
\usepackage{enumitem}
\usepackage{xspace}

\usepackage{caption}
\usepackage{listings}

\newcommand{\eg}{\textit{e.g.,}}
\newcommand{\ie}{\textit{i.e.,}}
\newcommand{\etal}{et al.}

\newcommand{\acronym}{\textsc{PromptPET}\xspace}
\newcommand{\decoy}{{Decoy}\xspace}

\newcommand{\ruleref}[1]{\hypertarget{rulesrc:R#1}{}\hyperlink{rule:R#1}{R#1}}
\newcommand{\ruleback}[1]{\hyperref[sec:rule-analysis]{$\hookleftarrow$}}

\lstdefinestyle{prompt}{
    basicstyle=\ttfamily\footnotesize,
    breaklines=true,
    breakatwhitespace=true,
    frame=single,
    columns=fullflexible,
    keepspaces=true,
    showstringspaces=false,
    breakindent=0pt,
    escapeinside={(*@}{@*)},
    captionpos=b
}

\begin{document}

\title{\acronym{}: Prompt Obfuscation Against Profiling by AI Agents}

\author{Ke Yang}
\affiliation{%
  \institution{University of California, Irvine}
  \city{Irvine}
  \state{CA}
  \country{USA}
}
\email{ke.yang@uci.edu}

\author{Olivia Figueira}
\affiliation{%
  \institution{University of California, Berkeley}
  \city{Berkeley}
  \state{CA}
  \country{USA}
}
\email{olivia.f@berkeley.edu}

\author{Umar Iqbal}
    \affiliation{
      \institution{Washington University in St. Louis}
      \city{St. Louis}
      \state{MO}
      \country{USA}
      }
\email{umar.iqbal@wustl.edu}

\author{Athina Markopoulou}
\affiliation{%
  \institution{University of California, Irvine}
  \city{Irvine}
  \state{CA}
  \country{USA}
}
\email{athina@uci.edu}

\thispagestyle{plain}
\pagestyle{plain}

\begin{abstract}

Privacy is an important challenge when users interact with AI agents, since users share sensitive information, explicitly or implicitly, and AI agents provider can use this information for user profiling. In this paper, we aim to protect users' privacy via a user-side approach
that obfuscates sensitive information in a user's prompt before it is sent, while preserving enough information to elicit a useful response from the chatbot. This approach faces an inherent tradeoff: what an AI agent needs to produce a personalized and detailed response is largely what it uses to profile the user. To that end, we consider, evaluate, and compare four different obfuscation actions, namely redact, abstract, replace, and a novel decoy that we introduce, which keeps the true information, hides it among plausible alternatives that the online AI agent answers separately, and retains only the answer to the true one.
First, we systematically evaluate each obfuscation action independently in terms of the privacy-utility tradeoff it achieves.
Second, we propose \acronym{}, which follows a decision rule set to select an obfuscation action for each sensitive unit based on the unit's context. The rule set is searched by an LLM-based optimizer that iteratively proposes and refines rules based on their measured privacy and utility. On real user queries, \acronym{} delivers $3.3\times$ more privacy per unit of utility cost than the strongest prior approach and reaches $1.4\times$ higher privacy. Its advantage holds at the user level, where the provider accumulates a profile across all of a user's queries.

\end{abstract}

\maketitle

\section{Introduction}

AI agents are rapidly becoming an everyday computing interface, with platforms such as ChatGPT \cite{openai_chatgpt}, Gemini \cite{google_gemini}, and Claude \cite{anthropic_claude} collectively serving over one billion users \cite{nytimes_chatbots_influencers_brands_2026}.
Users interact with these systems for both personal and professional tasks, such as drafting work documents, planning travel, managing finances, and even seeking medical \cite{yun2025online} or legal advice \cite{liu2026llm}.
Unlike conventional computing systems, users interface with AI agents through open-ended natural language, often describing their goals, constraints, preferences, and circumstances in detail.
LLM-based agents, in turn, produce personalized, context-aware responses that are explicitly shaped by the information users provide~\cite{mireshghallah2024trust}.
This shift in the interaction paradigm encourages users to disclose richer and more sensitive information than they would in a search box, including details about their employment, health, financial status \cite{mireshghallah2024trust, tran2025understanding}, {\em etc.}

Unfortunately, this paradigm shift also introduces substantial and still poorly understood privacy and security risks.
As AI agents rely on rich contextual input to provide useful responses, users may disclose sensitive personal information without fully understanding how that information is collected, retained, reused, or shared~\cite{tran2025understanding, gumuselawareness}.
In addition to the usual data collection and tracking practices in interactions with all online services, there are unique challenges for user privacy in interactions specifically with AI agents.
First, the sensitivity of personal information that users reveal directly to AI agents is much higher than identifiers or personally-identifiable information (PII) traditionally used for tracking and advertising in online services \cite{qia2026beyond}. Second, AI agents are equipped with strong capabilities to indirectly infer additional sensitive attributes \cite{staab2024beyond}. Third, profiling is intertwined with personalization and often necessary to achieve better utility from AI agents.
These issues combined with  the scale of adoption in numbers (\ie{} over a billion users) and scope (the all-in-one role these platforms increasingly play across different contexts), and the emergence of stateful agents that retain memory, further exacerbate the privacy risks.
Although agent providers often offer some data transparency and control mechanisms, these remain limited in general
and particularly inadequate in protecting users' privacy from  AI agents.

\begin{figure}[t]
    \centering
    \includegraphics[width=\columnwidth]{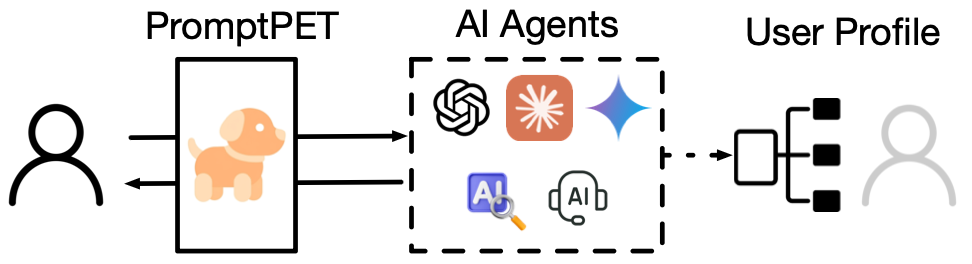}
    \caption{\textbf{\acronym{} Overview.} A user interacts with an AI agent through a conversational interface. \acronym{} sits between the user and the online AI agent, obfuscating sensitive parts of the user prompt (query) so that it subverts user profiling by the provider. See Figure \ref{fig:pipeline0} for details.}
    \label{fig:high-level}
\end{figure}

We argue that controlling the data shared in a prompt at the point of disclosure (\ie{} on the user side, before the data ever leaves the user's device and reaches the AI agent) is the most powerful privacy-enhancing technology (PET), as it does not rely on the provider and prevents downstream inferences. This approach is depicted in Figure~\ref{fig:high-level}.
\textit{Private computation}, such as cryptographic or trusted-execution techniques, also hides the prompt from the provider, but it must be deployed by the provider itself, and running the model elsewhere is not possible for closed-weight frontier models (Section~\ref{sec:related-work}).
A growing body of work that protects user prompt privacy at disclosure falls broadly into three areas: \emph{PII redaction}~\cite{chowdhury2025pr, de2026model}, \emph{data minimization}~\cite{zhou2025operationalizing, zhou2025rescriber, ngong2025protecting}, and \emph{obfuscation} operating on tokens and/or embeddings~\cite{tong2025inferdpt, wu2025cape}.
Prior work has focused primarily on protecting sensitive identifiers~\cite{chowdhury2025pr, zhou2025rescriber} and sensitive topics~\cite{chong2024casper}. It has not sufficiently considered the profiling of user interests, which are sensitive yet required to receive a useful response, or, more generally, profiling into fine-grained attribute taxonomies such as the IAB taxonomy used for advertising~\cite{iab_audience_taxonomy_1_1}.

In this paper, we focus on protecting the privacy of a user interacting  with an
AI agent, who profiles the user based on sensitive
information present in the user prompt.
To protect against that threat, we propose {\bf \acronym{}}, a principled and unifying privacy-utility optimized prompt obfuscation framework, which obfuscates user prompts, so as to subvert profiling attempted by the AI agent but also to preserve utility.
The idea is to optimize {\em which} sensitive units of information and {\em how} to modify them,  so as to control the privacy-utility tradeoff in principled way.
Our design of \acronym{} proceeds in two stages, making several methodological contributions along the way.

First, we consider, evaluate and compare four candidate obfuscation actions to apply to sensitive units of information within a prompt: {\em redaction, abstraction}, {\em replacement}, as well as a novel {\em decoy} action. While  false information injection has been used to enhance privacy in other domains, such as location privacy, anti-fingerprinting, obfuscating browsing data \cite{kido2005anonymous, shokri2012protecting, zhang2021harpo, adnauseam},
we design a scheme specifically for user prompts: it embeds each true unit among plausible decoys, instructs the online AI agent to answer each alternative separately, and retains only the answer to the true one, thereby preserving the utility of the response.
All four actions build on two design choices: (i) an importance score that estimates each sensitive unit's contribution to the response, which determines \emph{which} units to obfuscate and thereby controls the privacy-utility tradeoff, and (ii) a user-defined taxonomy of sensitive attributes, which defines \emph{what} is sensitive and guides \emph{how} each action transforms a unit.
We systematically compare the four obfuscation  strategies  in terms of their privacy-utility tradeoff, and we show that our proposed decoy outperforms the others: it achieves 45.3\% higher privacy than removal- and generalization-based actions.

Then, after evaluating the privacy-utility tradeoff of each single-action obfuscation strategy, we show that no single action is best for every sensitive unit: the right choice depends on the context. For example, a unit that is incidental in the query\footnote{In the rest of the paper, we refer to user ``prompt'', or user ``query'' interchangeably.} can be redacted at no cost to the response, whereas a unit that is the subject of the query must be kept and is better hidden among decoys.
Building on this intuition, we learn a \emph{decision rule set}, a short list of natural-language rules that map the context of a unit to the action that suits it. We formulate finding this rule set as a search problem and, following Yang \etal{}~\cite{yang2024large}, adapt an LLM as the optimizer, which iteratively proposes and refines rule sets based on their measured privacy and utility. At the single-query level, \acronym{} delivers $3.3\times$ more privacy per unit of utility cost than the strongest prior prompt obfuscation approach~\cite{ngong2025protecting, zhou2025operationalizing} and reaches $1.4\times$ higher privacy. We further evaluate \acronym{} at the user level, where the provider accumulates the profile across all of a user's queries. Although privacy drops over the first few queries, it stabilizes as the history grows, and \acronym{} retains its advantage over prior work, with $2.5\times$ more privacy per unit of utility cost and $1.3\times$ higher privacy.

The rest of the paper is organized as follows. Section~\ref{sec:related_work} discusses background and related work. Section~\ref{sec:threat_model} presents our setup and threat model. Section~\ref{sec:methodology} details the \acronym{} methodology. Section~\ref{sec:evaluation} presents our evaluation at both the single-query and user levels. Section~\ref{sec:discussion} discusses local-only answering, latency overhead, and stronger adversaries. Section~\ref{sec:conclusion} concludes the paper. We plan to release the prompts that comprise \acronym{} to enable future research.

\section{Background and  Related Work}\label{sec:related_work}

\subsection{Emergence of Stateful AI Agents}
AI agents are increasingly \emph{stateful}, retaining information from prior interactions in a persistent store, commonly referred to as memory.
This shift is now pervasive across a broad range of deployed systems, from general-purpose chatbots such as ChatGPT~\cite{openai_chatgpt}, Claude~\cite{anthropic_claude}, and Gemini~\cite{google_gemini}, to agentic browsers such as Atlas~\cite{openai_atlas} and Dia~\cite{dia_browser}, and general-purpose agents such as OpenClaw~\cite{openclaw} and Muse~\cite{meta_muse}.
At a high level, these systems retain information from past interactions and distill it into representations that may be useful in future ones~\cite{packer2023memgpt, chhikara2025mem0, park2023generative}.
Implementations vary in both what they retain and how they retain it.
Some summarize prior conversations~\cite{packer2023memgpt, zhong2024memorybank}, others extract salient entities and relationships~\cite{chhikara2025mem0, xu2026mem}, while others selectively retrieve historical interactions relevant to the current query~\cite{zhang2025survey}.
These mechanisms are already deployed at scale.
ChatGPT~\cite{openai_chatgpt}, for instance, persistently accumulates user information across sessions—both references to prior conversations and explicitly stored facts —to build an evolving profile of each user~\cite{openai_memory_controls_chatgpt}.
The key value provided by memory is that it enhances the utility of AI agents, allowing them to carry information across queries and sessions, rather than being constrained by the context window of the underlying LLM.
AI agents are also now increasingly transforming this accumulated state into an inferred profile of the user---their interests, preferences, goals, and beliefs---and leverage it to personalize future interactions~\cite{zhong2024memorybank, openai_memory_faq_2025, google_gemini_temporary_chats_privacy_controls}.
Importantly, memory is becoming a first-class feature of AI platforms.
Recent systems increasingly let users import, export, and transfer memory across applications~\cite{anthropic_import_memory, google_gemini_import_memory}, effectively turning accumulated user state into a portable asset rather than one confined to a single provider.

\subsection{Privacy Risks}
While statefulness makes AI agents more useful, it also exposes users to significant privacy risks, both \emph{manufactured} by the incentives of the platforms that operate the agents, and those \emph{intrinsic} to the agentic paradigm itself.
The manufactured risks stem from the fact that the personalization enabled by memory rests on continuous profiling of the user, and, as has repeatedly been the case in prior computing systems, such profiles are readily repurposed beyond improving the user's immediate experience~\cite{englehardt2016online,iqbal2023tracking, hannak2014measuring}.
Recent measurement studies show that user interactions are already routinely exposed to third-party services~\cite{wu2025depth, vekaria2025big}, and that this exposure occurs even when the LLM is merely mediating a user's interaction with a third party and has no functional need for the data itself~\cite{wu2025depth}.
Agents are also beginning to incorporate advertising and tracking services whose explicit purpose is to profile users~\cite{wu2025depth, jazlan2026tracking, nytimes_chatbots_influencers_brands_2026}, mirroring the surveillance-driven business models of earlier platforms---and, in some cases, drawing lawsuits over deceptive data practices~\cite{futurism_openai_personal_information_meta_google}.
Compounding this, providers exercise limited diligence over their ecosystems' privacy practices, some failing to implement even well-established privacy norms~\cite{wu2025depth}.

The intrinsic risk is more fundamental, i.e., to produce a useful response, the underlying LLM must receive the user's query in the clear.
Unlike prior computing platforms---for example, a web browser, whose vendor is not a party to what the user does on a page absent malicious behavior---in an AI agent disclosure is a precondition for utility, as the more a user reveals, the better the AI agent performs \cite{mireshghallah2024trust}.
Users thus have little choice but to trust the provider to be a faithful custodian of their data, yet they have no visibility into whether it honors its stated practices---and, as substantial prior work shows, vendors frequently do not~\cite{iqbal2023tracking, wu2025depth, andow2020actions, matte2020cookie, futurism_openai_personal_information_meta_google}.
Taken together, AI agents appear to be following the same profiling trajectory as prior platforms, but over a far richer, longitudinal data source and with no practical option for users to opt out\emph{---motivating a defense that protects the user before their data ever reaches the provider}.

\subsection{Related Work}
\label{sec:related-work}
Protecting user privacy in the agentic paradigm is subject to two non-trivial constraints.
First, a solution must strike a reasonable privacy-utility tradeoff, \ie{} it must sufficiently protect the user's sensitive information without degrading the usefulness of the system.
Second, it must operate without the cooperation of the vendor, which is often the very party against whom protection is needed.

Most closely related to our work is the growing body of work on prompt privacy, which falls broadly into the following five categories:
\emph{PII removal} substitutes named entities with placeholders before the prompt is shared with AI agents~\cite{chowdhury2025pr, chong2024casper, de2026model}, but cannot protect  non-PII sensitive information  that cannot be redacted.\footnote{We characterize, for the first time, the privacy threat of {\em AI-agent-side profiling} using non-PII information, such as interests. Using the IAB Audience Taxonomy~\cite{iab_audience_taxonomy_1_1}, we show that a profiler attacking PII-redacted prompts still recovers $1.93$ profile labels per query on average, showing that PII redaction is insufficient by design.} Consider the prompt \textit{``I am 35 years old, I was recently diagnosed with depression, and I live in Mountain View. Can you recommend a good therapist near me?''}.
While, the PII removal masks the age and location, it leaves the diagnosis and the request for a therapist untouched, which is revealing the user's mental-health status.
\emph{Data minimization} suppresses task-irrelevant context~\cite{zhou2025operationalizing, zhou2025rescriber, ngong2025protecting}, yet any information the user must disclose to obtain a useful response remains exposed by design.
In our running example, the therapist request cannot be removed without degrading utility.
\emph{Data anonymization} rewrites text to obscure privacy-sensitive content\cite{staab2024large,  yang2025robust}, but these methods optimize static text against an external reader, and their effect on the quality of an AI agents' responses —as opposed to text readability or a fixed classification label— is not evaluated.
\emph{Differential privacy} has been applied to user queries at the token or embedding level~\cite{tong2025inferdpt, wu2025cape}, which leaves sentence-level semantics exposed to profiling~\cite{mattern2022limits}, and at the sentence or document level~\cite{utpala2023locally}, which randomizes the entire query. Because this randomization is applied without regard to which specifics the response depends on, it sacrifices utility on the queries where those specifics are distorted.
Finally, \emph{Private computation} removes the provider's access to user query plaintext through cryptographic or trusted-execution techniques~\cite{hao2022iron}, but only if the provider chooses to deploy them, which users cannot enforce on today's frontier services. Users can avoid this dependence by running a model on their own device or switching to any provider that offers private inference, but because frontier models are closed-weight and can be served only by their providers, either alternative means falling back to a weaker model.

In contrast to these approaches, \acronym{} operates entirely on the user side and requires no cooperation from the AI Agent provider. It therefore fits the way AI agent services are delivered today, in which closed-weight frontier models are accessible only through their providers. It targets a query-level profiler that infers user attributes from the query as a whole, and therefore protects any sensitive information that supports such inference, including the information the task depends on, not only explicit PII or details that are irrelevant to the task. Rather than removing or randomizing the sensitive units that the response depends on, \acronym{} retains the important units on which the answer depends and conceals them among plausible decoys, while applying other obfuscation actions where they suffice. These actions are coordinated by rules that are learned to jointly optimize privacy against the profiler and the utility of the response (Section~\ref{sec:methodology}).

\section{Setup and Threat Model}
\label{sec:threat_model}
\textbf{System Model.}
Our setup is depicted in Figure \ref
{fig:high-level}.
We consider a user who interacts with an online AI agent through natural language conversation. Examples include general-purpose chatbots such as ChatGPT~\cite{openai_chatgpt} and Claude~\cite{anthropic_claude}, as well as LLM-backed search engines such as Google Search~\cite{google_homepage} and Bing~\cite{bing_homepage}.
A central part of the agent is an LLM that consumes the user's query and generates a response.
In practice, AI agents typically retain the user's interaction history and use it to build a user profile, which in turn serves to improve and personalize future responses.

\textbf{Adversary Capabilities and Goals.}
\label{sec:adversary}
We assume the AI agent's provider to be an ``honest but curious'' adversary. It sees and processes the user's queries in plain text, provides the correct responses, but also passively learns information in the process.
The adversary's goal is to profile the user, i.e., to infer and accumulate the user's interests and attributes from their queries over time.
In practice, such inferred profiles may be put to uses beyond delivering the requested functionality, such as targeted advertising or sharing with third parties.
We further assume the adversary organizes the profile according to a taxonomy of the kind used in practice.
In this paper, we adopt the IAB Audience Taxonomy~\cite{iab_audience_taxonomy_1_1}, as a representative example, which comprehensively captures user interests and is widely used for audience profiling in the advertising industry.

\begin{figure*}[!t]
    \centering
    \includegraphics[width=\textwidth]{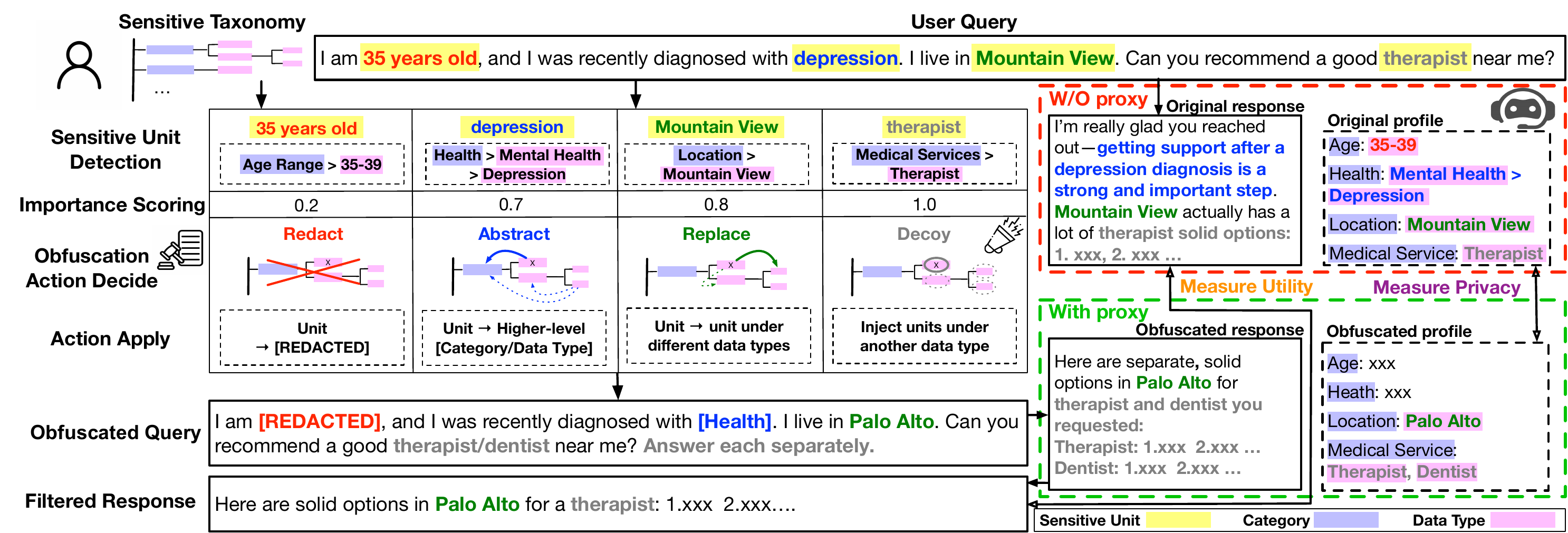}
    \caption{\textbf{Overview of the \acronym{} Framework.} For a given user query and sensitive taxonomy, \acronym{} detects sensitive units and estimates their importance scores, capturing each unit's contribution to generating the desired LLM response. An obfuscation action decider relying on a learned decision rule set then chooses one of four actions for each unit: redact, abstract, replace, or decoy. After all obfuscation actions are applied, the obfuscated query is sent to the online AI agent, which produces an obfuscated response and infers a user profile from the exposed query. The obfuscated response may contain answers to the decoys introduced by the decoy action, so \acronym{} filters them out and returns a filtered response to the user. The red dashed box shows the query flow without \acronym{}, whereas the green dashed box shows the flow with it. We evaluate utility by comparing the original and filtered responses, and privacy by comparing the original and obfuscated profiles. The two icons are cross-references: the megaphone icon on the decoy action and the gavel icon on the obfuscation action decider each reappear in a dedicated figure that details that component, namely Figures~\ref{fig:noise} and \ref{fig:rule-optimizer}, respectively.}
    \label{fig:pipeline0}
\end{figure*}

\textbf{Defense Goal and Scope.}
The goal of the defense is to obfuscate the query such that the adversary cannot reconstruct the user's sensitive attributes (in a given taxonomy),
while preserving the utility of the response, i.e., the user receives an answer as useful as if they had shared the full, un-obfuscated query.
The approach intercepts the user's query, applies the necessary obfuscation action, and only then forwards it to the online agent.
When the agent responds, the approach may intercept the response, apply complementary transformations (e.g., filtering content elicited by injected decoys), and return the answer to the user.
The approach runs on infrastructure the user controls, such as their own device or a cloud virtual machine they manage, and relies on an LLM hosted there to perform the obfuscation. The user trusts this infrastructure, the approach, and its LLM, none of which shares data with the adversary.

{\bf Out of Scope.}
A capable adversary may attempt to reverse-engineer the obfuscated information, for example, by observing the user's queries over time or by detecting contradictions across an account's obfuscated queries.
We do not consider de-obfuscation, and assume the adversary does not mount it.
Furthermore, we do not consider an adversarial AI agent that designs its responses to elicit sensitive information from the user.
While these are realistic attacks, they are different problems, deferred to future work.

\section{Methodology}\label{sec:methodology}

We propose \acronym{}, a user-side approach that optimizes the privacy and utility of online user-AI agent interactions. Achieving this goal requires addressing two challenges intrinsic to query obfuscation. First, the specifics of a query are precisely what make the response useful. The sensitive information it contains, which we formalize as \emph{sensitive units} in Section~\ref{sec:sensitive-unit-detection}, is often what allows the AI agent to tailor its answer to the user. Consequently, removing or altering these units protects privacy only at the cost of the answer's value. Second, no single obfuscation action is appropriate for every unit, and the most suitable action for a given unit depends on its context within the query.

In Section ~\ref{sec:noise}, we address the first challenge: we introduce \textbf{\decoy{}}, an obfuscation action that retains the original unit and supplements it with plausible but false decoys. A template combines them with the original query and instructs the online AI Agent to answer each alternative separately.
Upon receiving the response, the client filters out the decoy answers and returns only the original query's answer to the user.
Since the genuine specifics are remain in the sent query, the returned answer stays as personalized as the answer to the original query would be. But because the original unit appears alongside plausible decoys, a profiler observing the query would be distracted. In Section ~\ref{sec:rule-optimization}, we address the second challenge: we capture the dependence in a \textbf{decision rule set}, a short list of natural-language rules that map the context of a sensitive unit to the action that suits it. We formulate finding this rule set as a search problem and, following Yang \etal{}~\cite{yang2024large}, adapt an LLM as the optimizer, which iteratively proposes and refines decision rule sets based on their measured privacy and utility. In this way, \acronym{} learns to coordinate actions across diverse context, rather than relying on hand-crafted heuristics.

Next, we describe \acronym{} in detail, using Figure~\ref{fig:pipeline0} as a running example.

\subsection{Detecting Sensitive Information}
Our goal is to obfuscate the sensitive information in a user query, withholding the private content from the provider while avoiding a collapse in the utility of the AI agent's response. A prerequisite is to detect which unit leaks private information in the first place and, for each, how much it contributes to a useful response. The former delimits the scope of what should be obfuscated, while the latter guides how aggressively each unit can be obfuscated. \acronym{} therefore begins by detecting these units and scoring their importance, yielding the scope and the guidance on which the subsequent obfuscation step relies.
\subsubsection{Sensitive Taxonomy}\label{sec:sensitive-taxonomy}
The ``Sensitive Taxonomy'' specifies the scope of sensitive information that we aim to identify and protect within user queries.
Structurally, the taxonomy is a collection of trees, each corresponding to a category: the root node names the \textit{Category}, and the non-root nodes denote the \textit{Data types} within it. We define these terms more precisely, with examples, in Section~\ref{sec:sensitive-unit-detection}.
In this work, we adopt the IAB Audience Taxonomy~\cite{iab_audience_taxonomy_1_1}, a schema widely used in advertising and audience profiling, and we focus on its \textit{Demographics \& Interests} branches, which together contain 36 categories and 656 data types. In general, any taxonomy can be used in our methodology.

\subsubsection{Sensitive Unit Detection}\label{sec:sensitive-unit-detection}
Given a user query with a sensitive taxonomy, as shown in Figure \ref{fig:pipeline0}, \acronym{} first detects the following:
\begin{itemize}[leftmargin=*]
    \item \textit{Sensitive Units:} units in the query that disclose information that are useful for constructing the user profile within the scope of the sensitive taxonomy. For example, ``35 years old'' is a sensitive unit as it reveals the user’s age. Sensitive units are labeled with a category and data type from the sensitive taxonomy, as defined next. Sensitive units are highlighted in yellow in Figure \ref{fig:pipeline0}.
    \item \textit{Category:} the root node of the sensitive taxonomy that most relates to a sensitive unit. For example, ``35 years old'' maps to the \textit{Age Range}. Categories are highlighted in purple in Figure \ref{fig:pipeline0}.
    \item \textit{Data Type:} the specific attribute within a category in the sensitive taxonomy that most relates to a sensitive unit. Data types are finer-grained than categories and can be any node below the root node in the taxonomy. For example, ``35-39'' is a data type within the \textit{Age Range} category. A data type may also be hierarchical; for example, \textit{Depression} is a sub-data type of the data type \textit{Mental Health}, which belongs to the category \textit{Health}. Data types are highlighted in pink in Figure \ref{fig:pipeline0}.
\end{itemize}

This design allows \acronym{} to detect the sensitive information that the user wishes to protect and to map natural-language sensitive units to category/data types that can be used to guide the subsequent obfuscation.
Both BERT-based Named Entity Recognition (NER) models~\cite{devlin2019bert} and LLM-based methods~\cite{brown2020gpt3} are plausible choices for this task. We use an instruction-tuned LLM as our sensitive unit detector because it represents the current state-of-the-art for both locating and labeling sensitive information~\cite{garza2025prvl}.

\subsubsection{Importance Scoring} \label{sec:importance_scoring}
We observe that sensitive units may differ in how much they contribute to the AI agent's ability to produce the user's desired response.
Thus, capturing these differences is crucial for the downstream obfuscation actions.
For example, low-importance units may be obfuscated aggressively, whereas high-importance units require more utility-preserving treatment.
We therefore assign each sensitive unit a score in the range $[0,1]$, where higher values indicate greater importance.

We cast sensitive-unit importance scoring as a form of entity salience prediction~\cite{dunietz2014new}.
Since zero-shot instruction-tuned LLMs have been shown to achieve near-human performance on this task~\cite{bullough2024predicting}, we adapt this approach for our setting.
We validate this choice in Appendix~\ref{sec:importance-validation}.

\subsection{Obfuscation Actions}
\label{sec:obfucation_actions}
To obfuscate detected sensitive units, we consider four obfuscation actions: \textsc{Redact}, \textsc{Abstract}, \textsc{Replace}, and \textsc{\decoy{}} (Figure~\ref{fig:pipeline0}).
\textsc{Redact} and \textsc{Abstract} reduce the information a unit exposes, whereas \textsc{Replace} and \textsc{\decoy{}} introduce alternative or auxiliary information that weakens direct attribution and inference.
Among the four, only \textsc{\decoy{}} keeps the original unit intact. Overall, these actions provides \acronym{} with flexibility, enabling context-aware decisions.

\subsubsection{\textbf{Redact}}
The sensitive unit is removed and substituted with a placeholder token. In Figure~\ref{fig:pipeline0}, ``35 years old'' becomes \texttt{[REDACTED]}. As the most aggressive action, \textsc{Redact} eliminates the unit entirely while retaining its structural role in the sentence. Thus, it maximizes privacy gain per unit but discards all of the unit's task-relevant value, making it the least utility-preserving choice. This design follows prior data-minimization and masking methods~\cite{zhou2025operationalizing}.

\subsubsection{\textbf{Abstract}}
The sensitive unit is generalized to a higher-level node in the taxonomy. In Figure~\ref{fig:pipeline0}, ``depression'' becomes the coarser label \texttt{[Health]}. By mapping the unit to its parent category or data type, \textsc{Abstract} reduces the exposure of specific information while preserving enough context for the agent to respond appropriately, occupying a middle ground on the privacy--utility spectrum.

\subsubsection{\textbf{Replace}}
The sensitive unit is substituted with a plausible alternative drawn from a different data type under the same category. In Figure~\ref{fig:pipeline0}, ``Mountain View'' is replaced with ``Palo Alto.'' \textsc{Replace} preserves the structural and semantic role of the original unit while redirecting the profiler toward a misleading signal. Unlike \textsc{\decoy{}}, it \emph{discards} the true unit, so the information is lost.

\begin{figure}[t]
    \centering
    \includegraphics[width=\columnwidth]{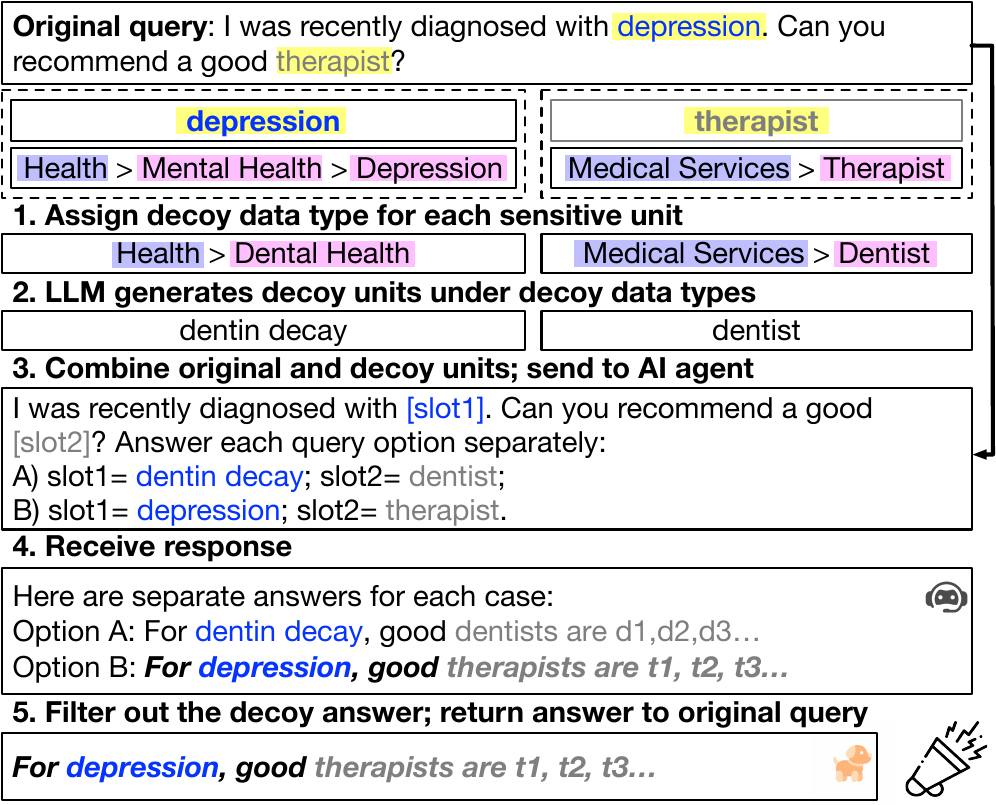}
    \caption{\textbf{Details of the \textsc{\decoy{}} Action.} (1) Each sensitive unit is assigned a decoy data type from the same category. (2) An LLM generates a decoy unit under each decoy data type. (3) A template combines the original and decoy units and asks the online AI agent to answer each option separately (details in Appendix~\ref{app:noise-template}). (4)--(5) The decoy answer is filtered out, and only the answer to the original query is returned to the user.}
    \label{fig:noise}
\end{figure}

\subsubsection{\textbf{\decoy{}}}
\label{sec:noise}
A key limitation of \textsc{Redact}, \textsc{Abstract}, and \textsc{Replace} is that they all degrade the original signal as they remove, coarsen, or overwrite the unit.
For high-utility contribution units, such obfuscations are suboptimal as they degrade utility.
Thus we propose \textsc{\decoy{}}, which retains the original unit and adds plausible information under a different data type.
The task-relevant content is left untouched, so the response is preserved, while making the original signal harder for a profiler to isolate, improving privacy without sacrificing utility.
In Section~\ref{sec:single-action}, we confirm that \textsc{\decoy{}} preserves utility on high-importance units, where removal-based actions instead cause a bimodal utility collapse.

Figure~\ref{fig:noise} details the \textsc{Decoy} action.
(1) For each unit, \acronym{} randomly selects a decoy data type that is irrelevant to the original unit but within the same category, here \textit{Health > Dental Health} and \textit{Medical Services > Dentist}. Irrelevance steers the profiler toward a false profile label, while staying within the category keeps the decoy plausible in context (\eg{} a \textit{Sports} decoy would be out of place in a health query). (2) The local LLM then generates a contextually plausible decoy unit under each selected decoy data type (``dentin decay'' and ``dentist''). The decoy units are generated jointly so that they stay consistent with each other in the query. (3) A template (Appendix~\ref{app:noise-template}) combines the decoy units with the original query: each sensitive unit is replaced by a placeholder, the decoy units (``dentin decay; dentist'') and the original units (``depression; therapist'') are listed as alternative filler options, and the online AI agent is instructed to answer each option separately. (4) The AI agent returns one answer per option, here dentist recommendations for dentin decay and therapist recommendations for depression. (5) \acronym{} discards the answer to the decoy units and returns only the original query's answer to the user.
Because the agent answers each option separately,
this filter can be hard-coded instead of requiring an extra local LLM call. On our test set, the filter correctly recovers the answer to the original query in $97.6\%$ of cases.

\subsection{Learning the Decision Rules}
\label{sec:rule-optimization}

In this section, we select the best obfuscation action, per sensitive unit. Specifically, given a query $x$ with $M$ detected sensitive units $u_1,\dots,u_M$, each labeled with a data type and an importance score, \acronym{} must assign one obfuscation action per unit, \ie{} choose an action combination $a \in \mathcal{A}^{M}$ with $\mathcal{A} = \{\textsc{Redact}, \textsc{Abstract}, \textsc{Replace}, \textsc{\decoy{}}\}$. Each action combination yields an obfuscated query whose privacy and utility we measure as in Section~\ref{sec:eval-setup}, and we summarize both as
\begin{equation}
    J_\alpha(x, a) = \alpha\cdot\mathrm{Privacy}(x,a) + (1-\alpha)\cdot\mathrm{Utility}(x,a),
    \label{eq:combined}
\end{equation}

where the weight $\alpha$ prioritizes privacy vs. utility. The four actions have complementary strengths (Section~\ref{sec:single-action}): no single action is best for every unit, and the right choice depends on the unit's importance, its data type, and the task type of the query. The per-query optimum $a^{\ast}(x) = \arg\max_{a} J_\alpha(x,a)$ is therefore the natural target. However, \acronym{} cannot search for it when a query arrives for two reasons: first, scoring a candidate action combination requires the online AI Agent's response to the corresponding obfuscated query, and a user-side tool can obtain that response only by sending the query, at which point the disclosure has already happened. Second, the number of candidates grows as $4^{M}$, so trying them all would be impractical even if it were safe.

\begin{figure}[t]
    \centering
    \includegraphics[width=\columnwidth]{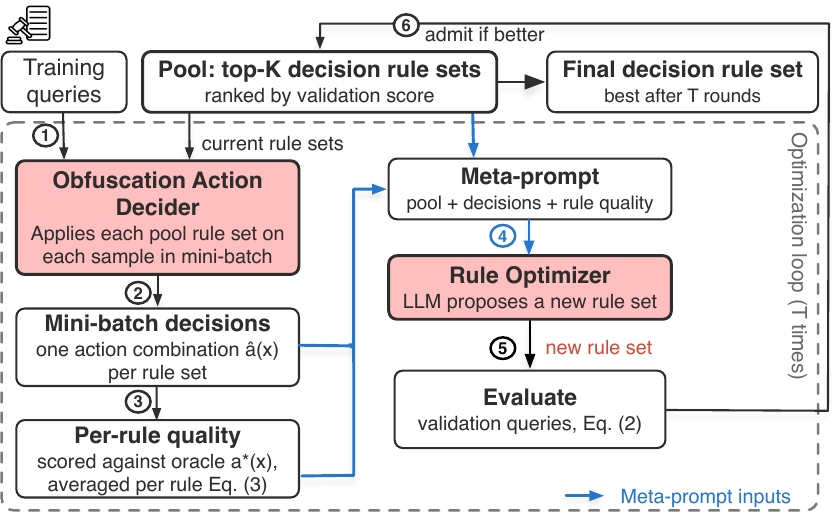}
    \caption{\textbf{Overview of the Rule-Optimization Framework.} Two LLM agents (shaded) cooperate in an iterative loop: the \emph{Obfuscation Action Decider} applies the top-$K$ rule sets to a training mini-batch, and the \emph{Rule Optimizer} proposes a new rule set from the meta-prompt, which joins the top-$K$ pool if it scores well on validation. Section~\ref{sec:rule-optimization} details steps 1--6.}
    \label{fig:rule-optimizer}
\end{figure}

Therefore, the choice must be made using only what is known before sending the query: each unit's data type and importance score, and the query itself. What \acronym{} needs is therefore a policy that maps this context to an action, learned offline on queries whose outcomes we can afford to measure. We represent this policy as a \emph{decision rule set} $\mathcal{R}$: a short, numbered list of natural-language rules of the form ``if $\langle$condition on the unit and the query$\rangle$, assign $\langle$action$\rangle$.'' Natural language rules can be followed directly by a local LLM without fine-tuning, and they remain readable to the user who deploys them. At runtime, the local LLM, the \textbf{Obfuscation Action Decider}, receives the query, its sensitive units with data types and importance scores, and $\mathcal{R}$, and outputs an action combination $\hat{a}_{\mathcal{R}}(x)$. The rule set is the only learned component, and we seek the one that performs best on a held-out validation set:
\begin{equation}
    \mathcal{R}^{\ast} = \arg\max_{\mathcal{R}}\;(
    \frac{1}{|\mathcal{D}_{\mathrm{val}}|}\sum_{x \in \mathcal{D}_{\mathrm{val}}} J_\alpha\big(x, \hat{a}_{\mathcal{R}}(x)\big)
    \;-\; \lambda\,|\mathcal{R}|),
    \label{eq:opt-target}
\end{equation}
where $\mathcal{D}_{\mathrm{val}}$ is a validation set, $|\mathcal{R}|$ is the word count of the rule set, and $\lambda = 10^{-5}$ penalizes verbose or redundant rules, keeping the learned policy compact and readable by the user who deploys it.

Eq.~\eqref{eq:opt-target} is a hard search problem: the solution space is discrete text, and the objective is a black box that involves the online AI Agent and scoring pipeline. Hand-writing $\mathcal{R}$ skips the search but encodes the author's bias rather than measured outcomes. We therefore follow Optimization by PROmpting (OPRO)~\cite{yang2024large}, in which an optimizer LLM is prompted with a \emph{meta-prompt} that contains a description of the optimization task and the top-$k$ solutions evaluated so far. At every step, the optimizer LLM proposes a new solution, which is scored and, if it ranks among the best, replaces the weakest member of the top-$k$ pool. Iterating this loop steadily improves the solutions. In our case, the solution is the decision rule set $\mathcal{R}$; the score is the validation objective of Eq.~\eqref{eq:opt-target}. The \textbf{Obfuscation Action Decider}, together with the scoring pipeline, plays the role of OPRO's scorer, and a second agent, the \textbf{Rule Optimizer}, plays the role of the optimizer LLM. OPRO's optimizer sees only (solution, score) pairs, which cannot reveal \emph{which} rule led to a poor decision. We therefore represent the solution as a numbered list of rules and give the optimizer feedback on each rule, measured against a per-query \emph{oracle}, the best action combination recorded for that query. Appendix~\ref{app:opro} explains why these adaptations are needed.

\textbf{Optimization loop.}
The framework maintains a pool of the top-$K$ rule sets ($K{=}5$), ranked by the objective in Eq.~\eqref{eq:opt-target} and seeded with an initial rule set (Section~\ref{sec:adaptive}). Each round then follows the numbered steps in Figure~\ref{fig:rule-optimizer}. (1) A mini-batch of $b{=}50$ training queries is sampled. (2) For every rule set $\mathcal{R}_i$ in the pool ($i = 1,\dots,K$) and every query $x_j$ in the batch ($j = 1,\dots,b$), the \textbf{Obfuscation Action Decider} produces the action combination $\hat{a}_{\mathcal{R}_i}(x_j)$ and reports the rule it relied on for each decision. (3) Each $\hat{a}_{\mathcal{R}_i}(x_j)$ is scored against the recorded oracle\footnote{Scores are pre-computed before optimization for only a subset of each query's $4^{M}$ action combinations.} $a^{\ast}(x_j)$, and the result is summarized per rule: for a rule $r$ of $\mathcal{R}_i$ used on the queries $\mathcal{C}_r$,
\begin{equation}
    q(r) = \frac{1}{|\mathcal{C}_r|} \sum_{x_j \in \mathcal{C}_r}
    \min\!\left\{1,\;
    \frac{J_\alpha\big(x_j,\hat{a}_{\mathcal{R}_i}(x_j)\big)}{J_\alpha\big(x_j,a^{\ast}(x_j)\big)}\right\},
    \label{eq:rule-quality}
\end{equation}
so that values near $1$ mark rules that reliably lead to a near-best combination and low values flag rules to revise or remove. (4) The meta-prompt is assembled from the rule sets $\mathcal{R}_1,\dots,\mathcal{R}_K$ with their validation scores, the per-rule quality $q(r)$, and each query $x_j$ paired with the decisions $\hat{a}_{\mathcal{R}_1}(x_j),\dots,\hat{a}_{\mathcal{R}_K}(x_j)$, so the optimizer can see where rule sets diverge on the same input. (5) The \textbf{Rule Optimizer} proposes a new rule set by keeping, revising, removing, or adding rules. (6) The new rule set is scored on $\mathcal{D}_{\mathrm{val}}$ using Eq.~\eqref{eq:opt-target} and is admitted if it outscores a pool member. After $T$ rounds, \acronym{} deploys the \textbf{Obfuscation Action Decider} with the best-scoring rule set in the pool, which approximates $\mathcal{R}^{\ast}$.

\section{Evaluation}\label{sec:evaluation}

Sections~\ref{sec:eval-setup} and~\ref{sec:experiment_setup} present the metrics and setup. In the single-query setting, Section~\ref{sec:single-action} characterizes each obfuscation action's privacy-utility tradeoff, Section~\ref{sec:adaptive} configures rule optimization, and Section~\ref{sec:main-comparison} evaluates \acronym{} against prior work. Section~\ref{sec:user-level} extends the evaluation to the user level.

\subsection{Evaluation Metrics}
\label{sec:eval-setup}
Evaluating a privacy-preserving approach requires measuring both \textit{privacy} (how much sensitive profiling signal is suppressed) and \textit{utility} (how well the chatbot’s helpfulness is preserved).

\subsubsection{Quantifying Privacy} \label{sec:privacy_gain}
As discussed in Section~\ref{sec:adversary}, the adversary seeks to infer a taxonomy-based user profile. We therefore quantify the
{\bf privacy gain} by comparing the original profile (\ie{} inferred from the original query) and the obfuscated profile (\ie{} inferred from the obfuscated query). Both profiles are expressed in the \textit{Demographics \& Interests} branch of the IAB Audience Taxonomy~\cite{iab_audience_taxonomy_1_1}.
Given the large size of our taxonomy, we improve stability by decomposing profiling into two stages: category extraction followed by data-type extraction within the extracted category.
To distinguish the inherent instability of the LLM-based profiler from the obfuscation-induced deviation, we evaluate self-consistency by running the profiler three times on the same query set and averaging the pairwise similarities between runs, resulting in $90.25\%$ Jaccard similarity, rising to $96.10\%$ when semantic similarity is considered.
For each original profile label, we compute the average embedding of its linked obfuscated profile labels and measure the normalized cosine similarity with it.
(We provide the details of similarity measurement in Appendix \ref{app:privacy-eva}.)

\subsubsection{Quantifying Utility}
\label{sec:utility_metrics}
We quantify {\bf utility retention} by comparing the filtered response with the original response generated directly from the original query. We evaluate utility from two complementary perspectives.
First, we compute the cosine similarity between the sentence embeddings \cite{reimers2019sbert} of two responses \cite{pape2025prompt, dong2025depth}.
This measures the coarse-grained semantic similarity between them.
Second, we evaluate fine-grained knowledge preservation.
This objective is closely related to summarization evaluation, which also measures information retention in generated text \cite{lin2004rouge}. We therefore adopt a claim-based evaluation strategy following \cite{scire2024fenice}, which assesses summary quality by extracting atomic knowledge claims from the source and using Natural Language Inference (NLI) \cite{williams2018broad} to determine whether those claims are supported by the summary.
Analogously, we extract atomic claims from the original response and use NLI to evaluate whether each claim is preserved in the filtered response. To better reflect our setting, we adapt the NLI labels to three categories: \textit{Present}, \textit{Abstracted}, and \textit{Absent}, corresponding to scores of $1$, $0.5$, and $0$, respectively. Intuitively, \textit{Present} indicates that the claim is preserved faithfully, \textit{Abstracted} indicates that it is retained only in a generalized form, and \textit{Absent} indicates that it is not preserved. Following \cite{metropolitansky2025towards}, which proposes a pipeline for extracting and verifying atomic factual claims from long-form text, we build the claim extractor by instruction-tuning an LLM and construct the NLI evaluator within the same framework to assess whether the extracted claims are preserved.
Overall, these two metrics capture both coarse-grained semantic similarity and fine-grained knowledge that remains. We weight them equally rather than rely on either alone for the utility evaluation.

\subsection{Experimental Setup}
\label{sec:experiment_setup}
\textbf{Dataset.}
Our dataset is derived from WildChat-1M~\cite{zhao2024wildchat}, a corpus of real user-chatbot interactions released after PII redaction.
For the single-query evaluation, we select queries that contain sensitive information within our scope, so that every test query gives the obfuscation actions something to act on.
We therefore build on Trust-No-Bot~\cite{mireshghallah2024trust}, which uses GPT-4~\cite{openai_gpt4} to extract conversations from WildChat-1M that contain sensitive information, yielding $3{,}319$ unique dialogues.
From each conversation, we retain only the first-turn query so that each query can be obfuscated and evaluated on its own.
We then apply our sensitive unit detector to identify information corresponding to our default sensitive taxonomy (Section~\ref{sec:sensitive-taxonomy}), extracting $7{,}911$ sensitive units from $2{,}182$ queries. These $2{,}182$ queries form the test set for all single-query experiments (Sections~\ref{sec:single-action}--\ref{sec:main-comparison}). Because this selection keeps only queries with in-scope sensitive information, it does not reflect a user's full query history. We therefore evaluate \acronym{} separately on the complete, unfiltered histories of real users in Section~\ref{sec:user-level}.
Our usage follows WildChat-1M's release license.

\textbf{User-side Models.} All LLM modules internal to \acronym{} use Gemma3:12B~\cite{deepmind_gemma3}. We use the \texttt{gemma3:12b} build distributed by Ollama~\cite{ollama} (4-bit quantized, $Q4\_K\_M$)~\cite{ollama_gemma3}, which occupies approximately 9\,GB of GPU memory at inference time. We choose Gemma3:12B because it is open-weight and fits on a single consumer-grade GPU, a hardware requirement that is acceptable for local deployment on the user side of \acronym{}.

\textbf{Adversary and Evaluator.} To approximate the adversarial chatbot provider and to measure privacy and utility, we use GPT-4.1 via the OpenAI API~\cite{openai_gpt41_models_doc}. GPT-4.1 stands in for a frontier production chatbot on the adversary side and serves on the measurement side.

\textbf{Reproducibility.} We set the temperature and random seed to 0 for all models.

\subsection{Single-Action Strategies Analysis}
\label{sec:single-action}
In this section, we analyze each obfuscation action in isolation to characterize its privacy-utility tradeoff. To control how much of a query is obfuscated, we apply an action only to the sensitive units whose importance score is at most a threshold $\tau$, so a larger $\tau$ obfuscates more units, including more important ones. We set $\tau \in \{0.6, 0.8, 1.0\}$ because $0.6$ and $0.8$ split the sensitive units into three groups of roughly equal size by importance score. Appendix~\ref{sec:importance-validation} gives the score distribution and validates the importance score itself. We evaluate each obfuscation action (\textsc{Redact}, \textsc{Abstract}, \textsc{Replace}, \textsc{\decoy{}}) at these thresholds, together with a no-obfuscation baseline $\tau = 0.0$, in which the original query is passed through unchanged. For every $(\text{action}, \tau)$ configuration, we score privacy and utility on all queries in our dataset, reporting utility as the mean of our two complementary metrics.

We examine the full per-query distribution of each configuration (Figure~\ref{fig:per-query-kde}); Table~\ref{tab:region-fractions} in Appendix~\ref{app:region-fractions} summarizes it numerically. To translate the distribution into interpretable statistics, we partition the unit square into four regions by splitting privacy and utility at their midpoint, $0.5$.\footnote{The split point is a reporting choice, not a parameter of the method. We use ($0.5$, $0.5$) for illustration, and the qualitative comparison between actions is unchanged under other reasonable cutoffs. A deployer may set each axis to the privacy and utility levels their application requires.} The regions are \emph{success} (high privacy, high utility), \emph{utility-collapse} (high privacy, low utility), \emph{privacy-miss} (low privacy, high utility), and \emph{failure} (low privacy, low utility). This setting will also be used in Section \ref{sec:main-comparison}, \acronym{} evaluation.

\begin{figure}[!t]
    \centering
    \includegraphics[width=\linewidth]{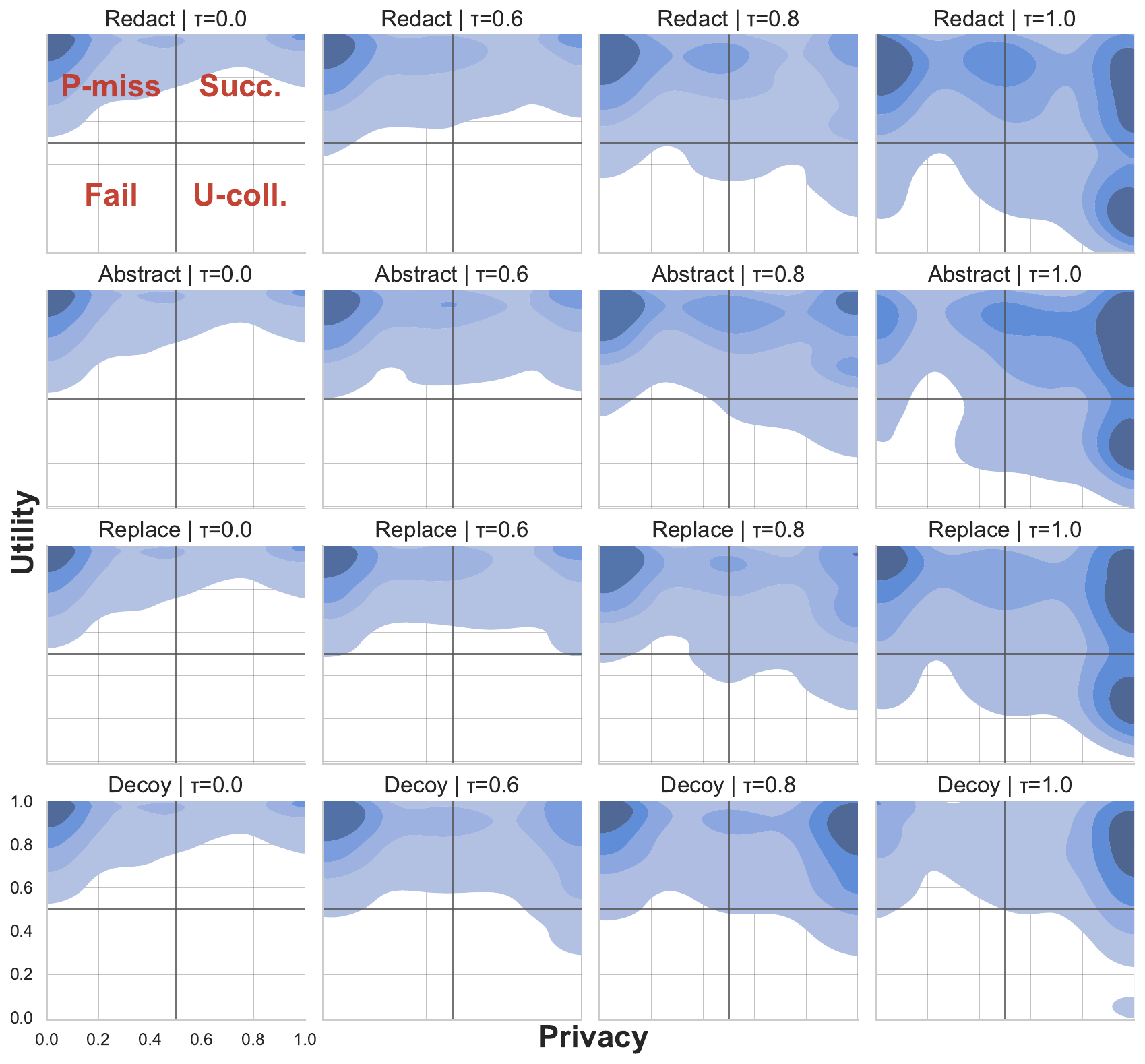}
    \caption{\textbf{Privacy-Utility Distributions.} Per-query distributions for each obfuscation action (rows) across different thresholds $\tau$. Each panel is a filled KDE over all queries. Crosshairs partition each panel into four outcome regions. A density floor ($0.05$) omits the $5.3\%$ lowest-density queries.
    }
    \label{fig:per-query-kde}
\end{figure}

\subsubsection{Per-Query Distribution}
\label{sec:per-query-distribution}
Figure~\ref{fig:per-query-kde} reports the joint per-query distribution of privacy and utility for each configuration as a 2D KDE (kernel density estimate), revealing how the query population traverses the privacy-utility space as obfuscation strength increases. Two features of the baseline panel ($\tau = 0.0$) merit comment. First, the no-obfuscation baseline does not collapse to the corner $(\text{privacy}, \text{utility}) = (0, 1)$: mean utility is $0.90$ rather than $1.0$, reflecting the inherent stochasticity of LLM-generated responses, and a small amount of mass appears at high privacy because $11.09\%$ of queries produce an empty original profile, meaning they contain no profilable information and therefore receive a privacy score of $1.0$. Second, residual privacy variance also arises from the internal instability of the profiler itself (Section~\ref{sec:privacy_gain}). As a result, $78.0\%$ of queries fall in the privacy-miss region at $\tau = 0.0$ (Table~\ref{tab:region-fractions}), confirming that, absent obfuscation, most queries leak their profile while still producing high-fidelity responses.

Reading each row from left to right (increasing $\tau$), all four actions share a common trend: the baseline mass concentrated in the upper-left corner spreads and migrates rightward, and the \emph{success} fraction grows monotonically from $21.1\%$ at $\tau = 0.0$ to between $39.4\%$ (\textsc{Redact}) and $73.7\%$ (\textsc{\decoy{}}) at $\tau = 1.0$. The \emph{manner} of migration differs sharply, however. \textsc{Redact} at $\tau = 1.0$ splits into three substantial modes: a dominant upper-right \emph{success} cluster of queries whose redacted units were inessential ($39.4\%$), a persistent upper-left \emph{privacy-miss} cluster ($31.2\%$), and a lower-right \emph{utility-collapse} cluster whose responses degraded once key units were deleted ($25.2\%$). The surviving privacy-miss mass indicates that our lightweight user-side detector misses many of the weak or implicit signals that the provider-side profiler still exploits. \textsc{Replace} behaves almost identically ($41.9\%$ success, $24.2\%$ utility-collapse, $29.6\%$ privacy-miss). \textsc{Abstract} exhibits a milder version of the same structure, as it converts much of the privacy-miss mass into successes (success $51.2\%$ vs.\ $39.4\%$ for \textsc{Redact}, while privacy-miss $20.1\%$ vs.\ $31.2\%$) at essentially no additional utility-collapse cost.

\textsc{\decoy{}}, by contrast, remains largely unimodal and migrates as a single coherent cluster toward the upper-right. At $\tau = 1.0$, $73.7\%$ of queries land in the success region, which is nearly double the fraction of \textsc{Redact} ($39.4\%$) and \textsc{Replace} ($41.9\%$). Meanwhile, only $7.4\%$ fall into the utility-collapse region, which is roughly a third of the $24$–$25\%$ observed for the other three. Because \textsc{\decoy{}} perturbs rather than deletes task-critical units, it preserves the surrounding response context. Its comparatively light privacy-miss mass ($17.9\%$) further suggests that perturbing a detected unit also disrupts the contextual cues the profiler would otherwise leverage, partially offsetting the detector's limited recall.

Finally, the failure region is sparse in every panel, with $0.5\%$ of queries at $\tau = 0.0$ and at most $4.3\%$ at $\tau = 1.0$, largely below the rendered density threshold. This is the area in which a query loses both privacy and utility, and its near-absence indicates that none of the actions sacrifice utility without a corresponding privacy gain, the basic property any sound obfuscation method should satisfy.

\begin{figure}[!t]
    \centering
    \includegraphics[width=\linewidth]{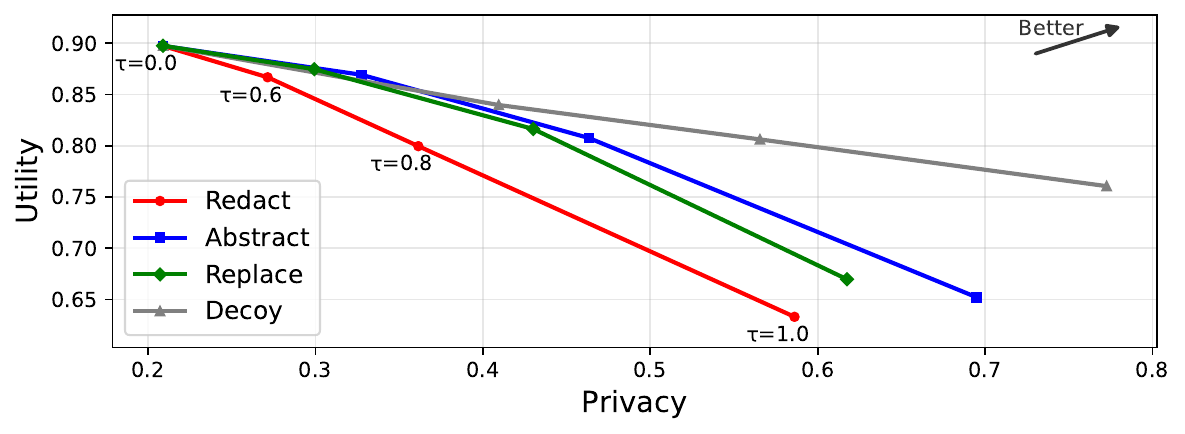}
    \caption{\textbf{Aggregate Privacy-Utility Tradeoff Curve.} Each curve traces an action (\textsc{Redact}, \textsc{Abstract}, \textsc{Replace}, \textsc{\decoy{}}).
    The arrow marks the preferred direction.}
    \label{fig:trade-off}
\end{figure}

\subsubsection{Aggregate Tradeoff Curve}
While Figure~\ref{fig:per-query-kde} shows the full per-query distribution,
Figure~\ref{fig:trade-off} summarizes each $(\text{action}, \tau)$ configuration by its \emph{mean} privacy and utility, tracing one tradeoff curve per action as $\tau$ sweeps from $0.0$ to $1.0$. All four curves originate from the shared baseline at $(\text{privacy}, \text{utility}) = (0.21, 0.90)$ and descend toward the lower-right as obfuscation strength increases, trading utility for privacy. A curve that lies up and to the right of another is strictly preferable. We emphasize that the curves are indexed by $\tau$, so two actions at the same $\tau$ generally sit at different privacy levels. A less aggressive action reaches a given $\tau$ at lower privacy and correspondingly higher utility. Thus, comparing actions at equal $\tau$ is misleading. The meaningful comparison is utility at the \emph{same privacy level}, the vertical separation between curves.

The four actions trace out clearly separated frontiers. \textsc{Redact} is at a disadvantage at every operating point: it incurs the steepest utility loss for a given privacy gain, reaching only $(0.59, 0.63)$ at $\tau = 1.0$. \textsc{Abstract} and \textsc{Replace} form a closely spaced middle tier: at the same privacy level, \textsc{Abstract} lies marginally above \textsc{Replace} throughout (\eg{} utility $0.81$ vs. $0.80$ at $\text{privacy} = 0.45$) and also extends to higher privacy ($0.69$ vs. $0.62$ at $\tau = 1.0$), making it the better of the two.
\textsc{\decoy{}} dominates the other three across most of the range. In the low-privacy area ($\text{privacy} \leq 0.37$), it sits marginally \emph{below} \textsc{Abstract} (\eg{} utility $0.863$ vs.\ $0.869$ at $\text{privacy} = 0.33$), so \textsc{Abstract} is the better choice under light obfuscation. Past the crossover near $\text{privacy} = 0.37$, however, \textsc{\decoy{}} separates decisively, and the gap widens monotonically: at $\text{privacy} = 0.55$, its utility is $0.81$, versus $0.75$ (\textsc{Abstract}), $0.72$ (\textsc{Replace}), and $0.66$ (\textsc{Redact}). It reaches the most favorable extreme point, $(0.77, 0.76)$, which is simultaneously the highest privacy and the highest utility of any action at $\tau = 1.0$. Over the full sweep, \textsc{\decoy{}} converts a $0.137$ utility loss into a $0.564$ privacy gain, a ratio of $4.11$, against $1.98$ (\textsc{Abstract}), $1.79$ (\textsc{Replace}), and $1.43$ (\textsc{Redact}).
This matches the per-query analysis: \textsc{\decoy{}} holds the population in the high-utility band while shifting it rightward, whereas removal-based actions buy privacy at a higher, more sharply rising utility cost. Yet the low-privacy crossover, where \textsc{Abstract} briefly overtakes \textsc{\decoy{}}, shows that the single best action is operating-point dependent. This is precisely the gap that motivates the \emph{selective} per-query strategy, discussed in Section~\ref{sec:adaptive}, in which each sensitive unit is assigned its optimal obfuscation action, and thus actions cooperate rather than compete with one another.

\subsubsection{Per-Query Optimal Action}
\label{sec:oracle}
We assess whether every action earns its place in the optimization pool: an action that is never the per-query optimum contributes nothing and could be dropped. Using a per-query oracle that, for each query and threshold, credits the action(s) maximizing $\alpha \cdot \text{privacy} + (1- \alpha) \cdot \text{utility}$ ($\alpha = 0.5$), with all tied actions credited (Table~\ref{tab:oracle-selection}). Every action is optimal for a non-trivial share of queries at every threshold. For example, even \textsc{Redact}, the weakest in aggregate, is the best on $11.0\%$ of queries at $\tau = 1.0$. Therefore, no action is dominated on a query-for-query basis and none can be dropped.
These niches are complementary and task-dependent. \textsc{Redact} is preferable when a sensitive unit is incidental to the query since deleting it costs no utility yet fully removes the attribute, whereas perturbation can leave it recoverable. \textsc{\decoy{}} is preferable when the unit is central to the query. For example, a search-style query asking where to obtain a specific title collapses under deletion, as the query loses its subject and returns a useless answer, whereas \textsc{\decoy{}} preserves the original query by adding a decoy alongside it and filters the decoy from the response.
The high tie rate at low $\tau$ ($62.4\%$ at $\tau = 0.6$) is mechanical: the importance filter marks few units, leaving most queries unchanged and the actions identical. As $\tau$ rises, the actions diverge, and a unique best action emerges for $95.6\%$ of queries at $\tau = 1.0$.
\begin{table}[!t]
    \centering
    \caption{\textbf{Optimal Action Frequency.} Percentage of queries for which each action is optimal under a
    per-query oracle.
    When actions tie, all tied actions are credited, so rows exceed $100\%$. Ties measure immaterial choices.}
    \label{tab:oracle-selection}
    \small
    \begin{tabular}{lccccc}
        \toprule
        $\boldsymbol{\tau}$ & \textsc{Redact} & \textsc{Abstract} & \textsc{Replace} & \textsc{\decoy{}} & Ties \\
        \midrule
        $0.6$ & $65.5\%$ & $70.2\%$ & $68.2\%$ & $78.0\%$ & $62.4\%$ \\
        $0.8$ & $41.1\%$ & $49.2\%$ & $47.9\%$ & $67.3\%$ & $36.9\%$ \\
        $1.0$ & $11.0\%$ & $21.0\%$ & $16.7\%$ & $58.3\%$ & $\phantom{0}4.4\%$ \\
        \bottomrule
    \end{tabular}
\end{table}

\subsection{Rule Optimization Configuration}
\label{sec:adaptive}
Using the methodology we developed in Section ~\ref{sec:rule-optimization}, we search a \emph{decision rule set} to help the obfuscation action decider assign each sensitive unit an optimal obfuscation action. From WildChat, we hold out a separate optimization set, disjoint from our test set, comprising $797$ training and $113$ validation queries. Using a mini-batch of $50$ queries, we run $20$ optimization rounds and retain the rule set that maximizes our optimization target (Eq.~\ref{eq:opt-target}) on the validation set. We first study how the seed decision rule set shapes the final result (Section~\ref{sec:seed}), then vary the weight $\alpha$ of the optimization target to prioritize privacy or utility (Section~\ref{sec:alpha}), and use the optimal configuration to evaluate \acronym{} in Section~\ref{sec:main-comparison}.

\subsubsection{Seed Decision Rule Set}
\label{sec:seed}
Because the starting point of a black-box search can determine which optimum it reaches, and OPRO is known to be sensitive to initialization~\cite{yang2024large}, we evaluate several seed rule sets derived from our findings (empty, threshold-based, and oracle-based) in Section~\ref{sec:adaptive}.
We optimize the decision rule set for the balanced objective ($\alpha = 0.5$) and study how the seed decision rule set influences the quality of the final decision rule set. We consider three seed decision rule sets of increasing prior knowledge: (i)~\textsc{Empty}, no rules; (ii)~\textsc{Threshold}: two threshold rules based on Figure~\ref{fig:trade-off} (\ie{} if importance $>0.8 \Rightarrow$ \textsc{Noise}, else \textsc{Abstract}); and (iii)~\textsc{Oracle}:
importance-binned action priors derived from the per-query oracle (Table~\ref{tab:oracle-selection}) (see Appendix \ref{app:oracle-seed}). For each setting, we feed the seed set (\emph{Default}) and the final set (\emph{Best}) to the obfuscation action decider and report its performance, evaluated on the test set at $\tau=1.0$.
Table~\ref{tab:rule-optimization} shows that the rule-optimization raises \emph{both} privacy and utility for all three seeds, increasing the balanced objective by $+0.078$ (\textsc{Empty}), $+0.011$ (\textsc{Threshold}), and $+0.009$ (\textsc{Oracle}). The size of the gain shrinks as the seed improves, since a better seed leaves less to recover. \textsc{Empty} starts as the worst ($0.670$) and climbs the most, with the rule-optimization closing a large, roughly balanced deficit on both axes ($+0.078$ privacy, $+0.078$ utility), yet $20$ rounds from a cold start still leave it last overall ($0.748$). \textsc{Threshold} begins near a local optimum and moves little. \textsc{Oracle} is already strong at initialization, and its gain is almost purely privacy ($+0.016$ privacy, $+0.001$ utility), yielding the best final rule set ($0.764$ balanced) and the highest privacy and utility of any setting ($0.775, 0.753$).
\begin{table}[!t]
    \centering
    \caption{\textbf{Seed Decision Rule Set Comparison.}  \emph{Default} is the seed. \emph{Best} is the optimized rule set. ``Bal.'' is the balanced objective. Best final operating point in bold (\ie{} \textsc{Oracle}).}
    \label{tab:rule-optimization}
    \small
    \begin{tabular}{ll ccc c}
        \toprule
        Seed & Stage & Priv. & Util. & Bal. & $\Delta$Bal. \\
        \midrule
        \multirow{2}{*}{\textsc{Empty}}
          & Default & $0.689$ & $0.652$ & $0.670$ & \\
          & Best    & $0.767$ & $0.730$ & $0.748$ & $+0.078$ \\
        \midrule
        \multirow{2}{*}{\textsc{Threshold}}
          & Default & $0.757$ & $0.728$ & $0.743$ & \\
          & Best    & $0.760$ & $0.746$ & $0.753$ & $+0.011$ \\
        \midrule
        \multirow{2}{*}{\textsc{Oracle}}
          & Default & $0.759$ & $0.752$ & $0.755$ & \\
          & Best    & $\mathbf{0.775}$ & $\mathbf{0.753}$ & $\mathbf{0.764}$ & $+0.009$ \\
        \bottomrule
    \end{tabular}
\end{table}

\subsubsection{Optimization Target}
\label{sec:alpha}
The optimization target $\alpha\cdot\text{privacy} + (1-\alpha)\cdot\text{utility}$ sets how strongly the optimizer favors privacy over utility. Starting from the \textsc{Oracle} seed, we optimize three targets, namely utility-only ($\alpha=0$), balanced ($\alpha=0.5$), and privacy-only ($\alpha=1$), reporting performance at $\tau=1.0$ on the test set.

The three settings occupy a tight region: privacy within [0.78, 0.79] and utility within [0.74, 0.76], which shows that $\alpha$ offer fine control over the operating point rather than a coarse tradeoff. Consistent with the objective, utility-only attains the highest utility ($0.758$) and privacy-only the highest privacy ($0.791$), but both margins are within $0.01$ of the balanced setting ($0.775$ privacy, $0.753$ utility). This narrow spread is \emph{expected} at $\tau = 1.0$: every detected unit is obfuscated regardless of $\alpha$, so the weight only reshapes \emph{which} action each unit receives, not whether it is protected.

The action mix confirms that $\alpha$ shifts the decision in the intuitive direction. Relative to the balanced default, leaning toward utility trades the lossy \textsc{Abstract} action almost entirely for response-preserving \textsc{\decoy{}} (\textsc{Abstract} $9.6\% \rightarrow 0.7\%$, \textsc{\decoy{}} $84.8\% \rightarrow 92.5\%$), whereas leaning toward privacy  raises the share of the destructive \textsc{Redact} action ($5.4\% \rightarrow 8.0\%$). The weight thus steers the rule set toward utility-preserving or privacy-maximizing actions as intended. But the effect on aggregate privacy and utility is small only because \textsc{\decoy{}} already scores well on both axes, leaving little to gain by substituting other actions. This also explains why strong privacy comes for free: even the utility-only objective reaches $0.782$ privacy, since $\tau = 1.0$ obfuscates every unit, and optimizing for utility prefers utility-preserving actions over destructive ones.

\subsection{\acronym{} Evaluation}
\label{sec:main-comparison}
\begin{figure}[!t]
    \centering
    \includegraphics[width=\columnwidth]{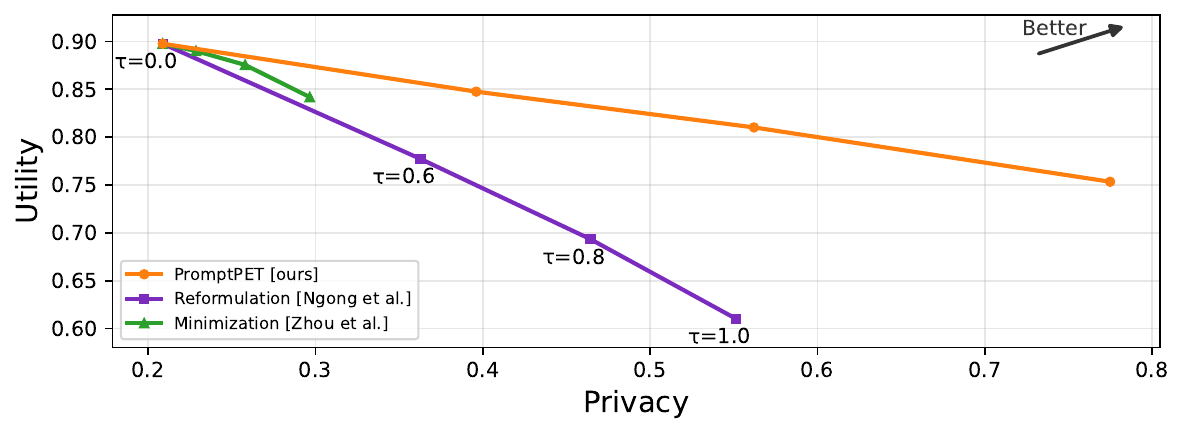}
    \caption{\textbf{\acronym{} Privacy-Utility Tradeoff Curve.} Includes \acronym{} (orange) and prior work.
    The arrow marks the preferred direction.
    }
    \label{fig:main_figure}
\end{figure}

Guided by Sections~\ref{sec:seed} and \ref{sec:alpha}, we fix \acronym{}'s default configuration as the \textsc{Oracle}-seeded rule set under the balanced objective ($\alpha=0.5$). We adopt the balanced objective as a preference-neutral default that favors neither axis; the utility- and privacy-only decision rule sets remain available for deployments that weigh one objective more heavily. The resulting decision rule set is listed in Appendix~\ref{app:rules}, and we analyze how individual rules shape the action decisions in Appendix~\ref{sec:rule-analysis}. We evaluate \acronym{} against prior automated defenses, sweeping the threshold $\tau \in \{0.0, 0.6, 0.8, 1.0\}$, and report the resulting tradeoff curve in Figure~\ref{fig:main_figure}.

\subsubsection{Baselines}
We compare against the two previous works that automate the full obfuscation pipeline without human intervention and operate on free-form rather than fixed-schema data, excluding work that targets a different adversary. The first is the contextual-reformulation framework of Ngong \etal{}~\cite{ngong2025protecting}, which rewrites out-of-context information in the prompt. The second is the zero-shot LLM minimizer of Zhou \etal{}~\cite{zhou2025operationalizing}, which retains, redacts, or abstracts each sensitive unit. However, no prior work targets a sensitive scope as broad as ours. To place both baselines within our scope, we fix the detection stage across them: each baseline is offered the same sensitive units that \acronym{} selects under the importance threshold $\tau$ (Section~\ref{sec:single-action}). This only restricts which units are offered to a baseline, and it remains free to retain a candidate it judges in-context or task-necessary, as in its design.

\subsubsection{Results}
Figure~\ref{fig:main_figure} shows that \acronym{} dominates both baselines by a wide margin. The method of Ngong \etal{} (purple)\cite{ngong2025protecting} degrades utility steeply per unit of privacy and saturates at (privacy, utility) = $(0.55,0.61)$. At the same privacy level, \acronym{} retains utility at $0.81$ and reaches a higher privacy of $0.77$.
The zero-shot minimizer by Zhou \etal{} (green)\cite{zhou2025operationalizing} is gentler but plateaus at a privacy level of $0.30$, only marginally above the no-obfuscation baseline.
This is consistent with the authors' own finding: a directly prompted LLM exhibits a bias toward abstraction that leads to oversharing rather than sufficient removal~\cite{zhou2025operationalizing}, leaving a residual signal that our adversarial profiler recovers. Their offline search would fare better but is not a deployable approach.
Quantitatively, \acronym{} delivers $3.3\times$ more privacy per unit of utility cost than the strongest prior baseline ($k = 3.9$ vs.\ $1.2$) while reaching $1.4\times$ higher privacy ($0.78$ vs.\ $0.55$), decisively outperforming prior work.

\subsubsection{Failure Analysis}
\label{sec:error-analysis}
To characterize where \acronym{} falls short, we partition its $\tau=1.0$ outcomes with the same region cut as Section~\ref{sec:per-query-distribution} (\ie{} privacy and utility split at $0.5$). \acronym{} reaches the \emph{success} region on $73.8\%$ of queries. The remainder splits into two error modes: under-protection (privacy-miss, $17.5\%$) and over-obfuscation (utility-collapse, $7.6\%$), with simultaneous failure of both axes being rare ($1.1\%$).
The larger error mode is privacy-miss, meaning queries whose profiles the adversary can still recover. These are not failures to \emph{act}, as every privacy-miss query had sensitive units detected and obfuscated (mean $3.1$ units), yet their privacy is near zero (median $0.00$, with $80\%$ below $0.2$). Because the detected units were obfuscated and the leak persisted, the adversary must reconstruct the profile from residual, co-occurring context rather than from the obfuscated units themselves. Consistent with this, privacy-miss queries rely less on \textsc{\decoy{}} and more on \textsc{Redact}/\textsc{Abstract} (\ie{} the actions least able to hide contextual cues) compared to the success region ($78\%$ vs.\ $86\%$ \textsc{\decoy{}}).
The smaller error mode, utility-collapse, stems from \textsc{Redact}: these queries apply it at more than twice the success-region rate (9\% vs. 4\% of unit), so the lost utility comes from deleting a unit that the answer needed, whereas \textsc{\decoy{}} seldom breaks a response. The number of obfuscated sensitive units is the same as in the success region (median 2), so the failure is not due to over-processing but rather applying a destructive action to an important sensitive unit.
The two modes bound the headroom of \acronym{} and point to distinct improvements. In particular, closing the under-protection gap calls for detecting implicit, contextual signals beyond the explicitly flagged sensitive units, while closing the over-obfuscation gap calls for more reliable routing of important sensitive units away from \textsc{Redact}.

\begin{figure}[!t]
    \centering
    \includegraphics[width=\columnwidth]{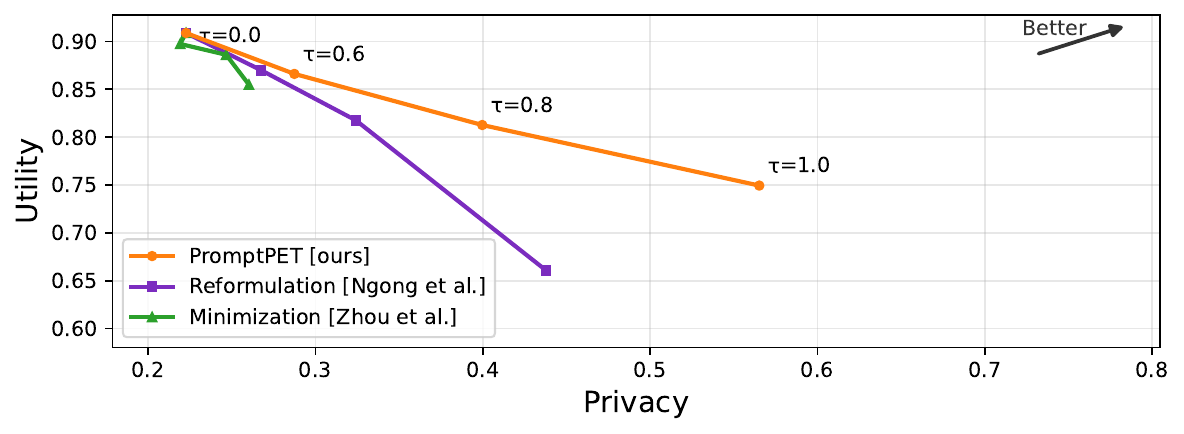}
    \caption{\textbf{User-Level Privacy-Utility Tradeoff.} Each point is the mean over 100 users of per-user privacy and per-user utility, for $\tau \in \{0.0, 0.6, 0.8, 1.0\}$. Axes match Figure~\ref{fig:main_figure}.}
    \label{fig:user-tradeoff}
\end{figure}

\subsection{User-Level Evaluation}
\label{sec:user-level}

Sections \ref{sec:single-action}--\ref{sec:main-comparison} evaluate queries independently, whereas a stateful provider can accumulate a user's profile across all their queries. In this section, we apply \acronym{} to all queries of each user and evaluate the privacy-utility tradeoff at the user level.

\subsubsection{Experimental Setup}
\label{sec:multiturn-setup}
We treat all queries that share a hashed IP address in WildChat as coming from one user. We group English-language users into five buckets by their total number of queries (1--5, \ldots, 21--25)\footnote{91.7\% of English WildChat users send at most 19 queries.} and sample 20 users from each bucket, yielding 100 users and 1,297 queries (details in Appendix~\ref{app:multiturn-buckets}). \acronym{} obfuscates each query independently, and the online AI Agent answers it in the context of its conversation, i.e., the earlier obfuscated queries and its own responses.

To measure privacy, we profile each of a user's queries with the same profiler as in the single-query evaluation, both before and after obfuscation, and pool the resulting labels into an original and an obfuscated user profile. We then compare the two pooled profiles with the same similarity-based measurement used for single queries (Section~\ref{sec:privacy_gain}, Appendix~\ref{app:privacy-eva}),
which  we refer to as \emph{per-user privacy}. We assess \emph{per-user utility} as the mean utility (Section~\ref{sec:utility_metrics}) over the user's queries. The baselines and the $\tau$ sweep follow Section~\ref{sec:main-comparison}.

\subsubsection{Results}
\label{sec:user-level-results}
Figure~\ref{fig:user-tradeoff} shows that \acronym{} keeps its advantage over both baselines at the user level. At $\tau=1.0$, it reaches a per-user privacy of $0.57$ at a utility of $0.75$, whereas Ngong \etal{}~\cite{ngong2025protecting} (purple) saturates at $(0.44, 0.66)$, and Zhou \etal{}~\cite{zhou2025operationalizing} (green) stays at a privacy of $0.26$, close to the no-obfuscation point ($0.22$). At matched utility ($\tau=0.8$, utility $0.81$ vs.\ $0.82$), \acronym{} still adds $0.08$ privacy over baselines. Measured as in Section~\ref{sec:main-comparison}, \acronym{} gains $2.5\times$ more privacy per unit of utility cost ($k = 2.1$ vs.\ $0.9$) while reaching $1.3\times$ higher privacy.

\begin{figure}[!t]
    \centering
    \includegraphics[width=\columnwidth]{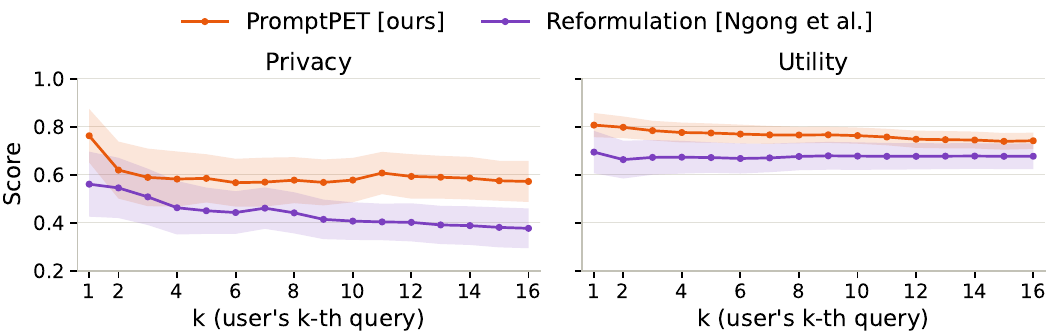}
    \caption{\textbf{Privacy and Utility as Queries Accumulate} ($\tau=1.0$). Per-user privacy (left) and utility (right) pooled over each user's first $k$ queries, averaged over the 40 users with at least 16 queries, comparing \acronym{} with the reformulation baseline. Shaded: 95\% CI of the mean.}
    \label{fig:accumulation-min16}
\end{figure}

\subsubsection{Privacy as Queries Accumulate}
\label{sec:user-level-history}
On the same query-set, the mean per-query privacy at $\tau=1.0$ is $0.72$, but per-user privacy is $0.57$, because any sensitive information leaked by even one query stays in the user's pooled profile. To see how this drop develops, Figure~\ref{fig:accumulation-min16} pools each user's first $k$ queries in time order. For \acronym{} (orange), privacy drops from $0.76$ after the first query to $0.59$ by the third, then stabilizes between $0.57$ and $0.61$ through query 16. From the third query on, even the lower end of the 95\% CI of the mean (shaded) stays above $0.46$. Utility declines only slightly, from $0.81$ to $0.74$. The reformulation baseline of Ngong \etal{}~\cite{ngong2025protecting} (purple)\footnote{We omit the minimization baseline of Zhou \etal{}~\cite{zhou2025operationalizing} from this figure: it obfuscates conservatively, leaving $78\%$ of queries unchanged, so its privacy stays close to the no-obfuscation level throughout ($0.26$ vs.\ $0.22$ at the user level).} behaves differently: its privacy starts lower, at $0.56$, and keeps falling over the whole history, reaching $0.38$ at query 16 with no sign of leveling off. This contrast is consistent with the role of \textsc{\decoy{}}. Reformulation only removes or generalizes information, as is done by \textsc{Redact} and \textsc{Abstract}. Such actions keep the current query from leaking but add nothing to the provider's view of the user, so every future leak lowers the user's privacy further. \textsc{\decoy{}} instead introduces false information, which stays in the provider's pooled profile and continues to dilute as the queries accumulate.

\subsubsection{Dependent Queries}
\label{sec:multiturn-dependent}
Unlike the single-query setting, a multi-query chat may mix new queries with follow-ups that build on earlier queries. A follow-up depends on the earlier context, and that context has been obfuscated. Obfuscation in one query can thus affect later queries, an effect that the single-query evaluation cannot capture. To measure it, we label every query after the first in each conversation as either \emph{independent}, a new request ($51\%$), or \emph{dependent}, a follow-up on an earlier response ($49\%$), and compare the two types. The labels are produced by GPT-5~\cite{openai_gpt5} with the annotator prompt of Huang \etal{}~\cite{huang2026llms} (Appendix~\ref{app:multiturn-labeling}). Dependent queries are harder for \acronym{} in utility but not in privacy. Their utility is $0.67$, versus $0.79$ for later independent queries, a statistically significant difference (two-sided Mann--Whitney $U$ test, p-value $<10^{-11}$). The gap also holds when each user's dependent queries are compared with that same user's independent queries (mean difference $-0.10$), so it is not caused by differences between users. Privacy on queries that reveal a profile, in contrast, is the same for both types ($0.56$ vs.\ $0.57$, p-value $=0.95$), and again within the same user (mean difference $-0.02$, p-value $=0.94$).

To locate the utility gap, we consider two ways obfuscation can affect a query: through the earlier queries of its conversation and through the query itself. The earlier queries are the main cause. When the query itself is sent unchanged, an obfuscated history lowers utility from $0.92$ to $0.71$ for dependent queries ($-0.21$), but only from $0.97$ to $0.84$ for independent ones ($-0.13$). A follow-up builds directly on the provider's earlier response, and that response was generated from an obfuscated query; \eg{} a request to ``make it shorter'' then shortens a response that already differs from the one the user would have received. Obfuscating the query itself costs both types about the same: on top of an obfuscated history, utility falls further to $0.59$ for dependent and $0.73$ for independent queries ($-0.12$ vs.\ $-0.11$). The gap therefore stems from how follow-ups inherit earlier obfuscation, not from how they are obfuscated themselves.
Addressing this calls for obfuscation that spans queries, \eg{} selectively bringing back earlier obfuscated units
when a follow-up depends on them, which we leave to future work.

\section{Discussion}
\label{sec:discussion}
\textbf{Local Models.}
\acronym{} uses a local LLM to obfuscate queries. A natural question is whether the local LLM could also answer the queries, so that they never leave the user side and privacy is fully preserved. We expect this to degrade utility, since answer quality would then depend on a much weaker model, and our measurements support this. Answering directly with \texttt{gemma3:12b}, the model \acronym{} uses, achieves overall utility close to \acronym{}'s ($0.729$ vs.\ $0.753$ at $\tau{=}1.0$), but the shortfall is concentrated in the tail: on the quarter of queries with the worst local answers, mean utility is only $0.51$. Routing the same queries through \acronym{} raises it to $0.628$ at $\tau{=}1.0$ (privacy $0.77$), $0.705$ at $\tau{=}0.8$ ($0.56$), and $0.745$ at $\tau{=}0.6$ ($0.40$), so users can choose the tradeoff they need. The gap widens when the user's hardware supports only a smaller model. Answering directly with \texttt{gemma3:4b} lowers overall utility to $0.692$ and tail utility to $0.466$, whereas \acronym{} with the 4B model still reaches $0.752$ overall at $\tau{=}1.0$, nearly the same as with 12B, and lifts the tail to $0.648$, at a privacy of $0.63$ instead of $0.77$. This is because in \acronym{} the local model only obfuscates while the cloud model answers, so a weaker local model costs some privacy but no utility. As cloud models advance and small open models improve, we expect this gap to grow while the hardware \acronym{} needs keeps shrinking.

\textbf{Latency Overhead.}
\label{sec:time-cost}
We measure the latency of \acronym{} on a single NVIDIA RTX A5000 and compare it with the provider's response time on the raw queries. Because the distribution is right-skewed, we report medians. \acronym{} adds $11.6$\,s of client-side latency per query, dominated by detection ($5.5$\,s). With the provider's response time ($4.7$\,s, without \acronym{}), the end-to-end median is $17.1$\,s. This is still faster than local-only answering with the same model ($22.4$\,s). We treat this overhead as real but secondary to the privacy-utility trade-off, and two factors bound its impact. First, the overhead is opt-in and per-query rather than a system-wide tax: a user enables obfuscation for sensitive prompts and bypasses it otherwise. Second, its dominant term is detection, which is unoptimized local inference and shrinks with a smaller model. With \texttt{gemma3:4b}, the client-side latency falls to $5.5$\,s (detection $3.2$\,s), and the end-to-end median falls to $10.2$\,s. This suggests that a smaller, purpose-built detector could reduce the overhead even further.

\textbf{Further Attacks.}
Our attack model (Section~\ref{sec:threat_model}) assumes an honest-but-curious provider that infers labels from each query independently and does not actively try to de-obfuscate. It can be extended in several ways. First, the attacker currently takes the union of all labels as the user profile; a variation would infer labels from an entire conversation. Second, a more capable adversary could target \decoy{} in two ways. (i) In a multi-query conversation, each query is obfuscated independently, so the decoy may change from query to query while the user's real thread stays consistent. An adversary comparing queries could thus tell the real thread from the decoy. A natural defense is to maintain the decoy thread across queries, keeping the same decoy data type and continuing it coherently so it looks like a second user rather than a one-off insertion. (ii) \decoy{} currently pairs each real unit with a single decoy, which bounds an adversary's re-identification rate from below at 50\%. Generating more, better-fitting decoys per unit would lower this floor and make the real unit harder to reveal. Evaluating both defenses against an adaptive adversary remains future work.

\section{Conclusion}
\label{sec:conclusion}
AI agents pose privacy risks to users, as the sensitive information in their queries can be used beyond answering them, such as for user profiling. In this paper, we develop and evaluate \acronym{}, a user-side approach that obfuscates sensitive information beyond PII in user queries at the time of disclosure, while optimizing the privacy-utility tradeoff.
\acronym{} uses four obfuscation actions: redact, abstract, replace, and a novel decoy action, which keeps the true information, hides it among plausible alternatives, and retains only the answer to the true one. We first evaluate each action in isolation and find that no single action is best for every sensitive unit. We therefore search for a decision rule set that guides the choice of action for each unit, using an LLM as the optimizer. On real user queries, \acronym{} delivers $3.3\times$ more privacy per unit of utility cost than prior state-of-the-art approaches and reaches $1.4\times$ higher privacy. Its advantage holds at the user level, where profiles accumulate across queries: privacy drops over the first few queries and then stabilizes. Two directions remain for future work: follow-up queries depending on earlier obfuscated turns, where utility is lower, and adaptive adversaries trying to tell the true information from the decoys. Overall, our results show that controlling disclosure at the source is a practical  and effective way to give users greater control over their privacy with AI agents. We plan to release the prompts that comprise \acronym{} to enable future research.

\section*{Acknowledgments}
This work was partially supported by the NSF Award 1956393 and 2543517 and a generous gift from the Samueli Foundation to the E+S Institute on AI (ProperAI) at UC Irvine. Olivia Figueira was supported by the UC Noyce Foundation, the UCI  ICS  Steckler Family Endowed Fellowship and the ARCS Danaher Foundation Scholar Award. We would like to thank Yu Duan from UCI for providing feedback on the figures of the paper.

\bibliographystyle{ACM-Reference-Format}
\balance
\bibliography{ref}

\appendix
\section{Importance Score Validation}
\label{sec:importance-validation}

\begin{figure}[h]
    \centering
    \includegraphics[width=\linewidth]{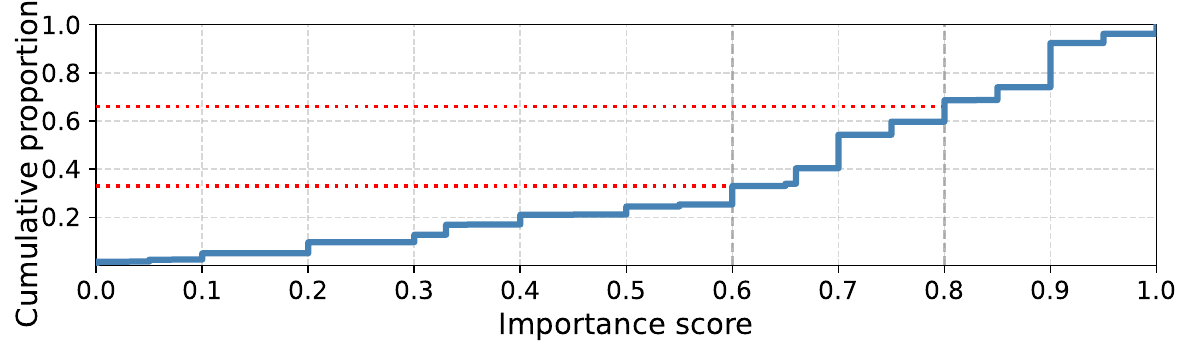}
    \caption{\textbf{CDF of Sensitive Unit Importance Scores.} Red dotted lines mark the $1/3$ and $2/3$ cumulative proportions.}
    \label{fig:importance_score}
\end{figure}

To characterize each obfuscation action's effect on the privacy-utility tradeoff independently, we select sensitive units for obfuscation according to the importance score $\mathcal{I}(\cdot)$, which estimates each unit's contribution to the response: an action is applied only to the units whose importance score is at most a threshold $\tau$ (Section~\ref{sec:single-action}).
By varying the threshold $\tau$, we control the set of units subjected to a given obfuscation action and, in turn, the operating point along the privacy-utility curve.
Before doing so, we must first validate the importance score itself, specifically whether it separates units that contribute substantially to the response.

\subsection{Importance Score Statistic}
\label{sec:importance-validation-statistic}
Figure~\ref{fig:importance_score} shows the empirical CDF of the importance scores assigned to these sensitive units, demonstrating the following two observations.
\emph{First, the distribution is right-skewed:} scores are concentrated toward the upper end of the range, which is intuitive for task-oriented queries, where real users tend to include information that specifies their requests rather than extraneous details. \emph{Second, equal-mass terciles are non-uniform in score:} the $1/3$ and $2/3$ cumulative proportions correspond to scores of $0.6$ and $0.8$, respectively, indicating that score-uniform thresholds (\eg{} $0.33$, $0.66$, $1.0$) would be suboptimal, as low thresholds cover relatively few units (units with score $\leq 0.33$ account for only $16.87\%$ of the total), while high thresholds traverse the dense upper region (units with score $\geq 0.9$ account for $25.94\%$), where threshold choice matters most for the privacy-utility tradeoff.

\begin{table}[h]
    \centering
    \small
    \setlength{\tabcolsep}{4pt}
    \caption{\textbf{Mean Utility Drop.} Utility drop when redacting equal number of sensitive units per importance score range.}
    \label{tab:importance-validation}
    \begin{tabular}{lccc}
        \toprule
        \textbf{\multirow{2}{*}{Importance range}}
            & \textbf{Low} & \textbf{Medium} & \textbf{High} \\
            & ($s \leq 0.6$) & ($0.6 < s \leq 0.8$) & ($s \ge 0.8$) \\
        \midrule
        Utility drop (cosine) & $0.0411$   & $0.0924$   & $0.1991$   \\
        Utility drop (NLI)    & $0.0918$   & $0.1588$   & $0.3380$   \\
        \bottomrule
    \end{tabular}
\end{table}

\subsection{Validation via Controlled Redaction} Before using the importance score to drive threshold selection, we verify that it actually distinguishes between high- and low-impact sensitive units. The most direct test is to redact the same number of units from each of three importance ranges and compare the resulting utility drops. If the importance score is meaningful, redacting high-importance sensitive units should degrade utility substantially more than redacting low-importance sensitive units.
We partition the $7{,}911$ sensitive units by score into three ranges: low ($s \leq 0.6$), medium ($0.6 < s \leq 0.8$), and high ($s > 0.8$). The initial counts are $2{,}610$, $2{,}813$, and $2{,}488$, respectively. To equalize range sizes, we randomly reassign boundary units (those with $s = 0.6$ or $s = 0.8$) to adjacent ranges, yielding exactly $2{,}637$ units in each range. For each range, we redact all of the detected sensitive units across the full dataset and measure the resulting utility drop per query, averaging over all queries.
Table~\ref{tab:importance-validation} reports the mean utility drop under two complementary metrics we defined in Section~\ref{sec:utility_metrics}: cosine similarity and NLI-based knowledge preservation. Both metrics exhibit a clear monotonic trend: redacting high-importance units causes a substantially larger utility drop than redacting medium- or low-importance units, roughly $4.84\times$ (cosine) and $3.68\times$ (NLI) for the drop incurred by redacting the same amount of low-importance units.

\section{\textbf{Privacy Evaluation}}
\label{app:privacy-eva}

To quantify the deviation between a original profile and its obfuscated profile, we have the following three observations:

\textbf{Observation 1: Exact-match comparison is insufficient.}
An obfuscated label may differ from the original label while still revealing substantial information if it is semantically close to the true one. For example, under the \textit{Age Range} category, replacing \textit{35-39} with \textit{40-44} provides much less protection than replacing it with \textit{75+}. Therefore, privacy gain should be measured by semantic similarity rather than by exact agreement alone.

\textbf{Observation 2: Semantic comparison is only meaningful within the same category.}
Labels from different categories describe different profile domains and should not be treated as relevant exposures. For example, an age label and an income label are not comparable, even if their embeddings happen to be accidentally close. Therefore, similarity should only be computed within the same category.

\textbf{Observation 3: Privacy depends on both residual exposure and injected uncertainty.}
When multiple obfuscated labels are associated with the same original label, labels close to the truth increase exposure, whereas distant or irrelevant labels increase ambiguity and can reduce exposure for that attribute. However, the successful hiding of one original label should not compensate for the strong leakage of another. Therefore, exposure should be aggregated separately for each original label and lower-bounded by zero before computing the profile-level score.

Guided by these observations, we define a profile exposure score as follows. For each original label, we identify all obfuscated labels linked to it within the same category (Observation 2), compute their mean embedding , and measure the normalized cosine similarity between this mean embedding and the embedding of the original label (Observation 1). Averaging embeddings before comparison naturally captures the interplay between residual exposure and injected uncertainty (Observation 3): obfuscated labels close to the truth pull the mean toward the original label and raise the score, while distant or irrelevant ones pull it away and lower the score. To prevent strong protection on one attribute from masking leakage on another, we lower-bound each per-label score at zero before averaging across labels to obtain the profile-level exposure score. We then define the privacy gain score as 1 - \text{exposure score}. We define the metric in detail as follows.

\textbf{Step 1}: Semantic similarity between taxonomy labels.
Each taxonomy label is represented as a \textit{category-data type} pair $(c_i,d_i)$, where $c_i$ denotes the category and $d_i$ denotes the specific data type. We encode each taxonomy label using an embedding model, denoted by $E(c_i,d_i)$. To reflect Observations 1 and 2, the similarity between two taxonomy labels $i$ and $j$ is defined as
\begin{equation}
\mathrm{Sim}(i,j)=
\begin{cases}
\cos\!\big(E(c_i,d_i), E(c_j,d_j)\big), & \text{if } c_i=c_j,\\[4pt]
0, & \text{otherwise.}
\end{cases}
\end{equation}

\textbf{Step 2}: Assign each obfuscated label to the original label it reveals most strongly.
Let the raw profile and the obfuscated profile be
\[
R=\{r_1,\dots,r_{N_R}\}, \qquad O=\{o_1,\dots,o_{N_O}\}.
\]
For each obfuscated label $o_i$, we identify the original label that it reveals most strongly:
\begin{equation}
r_i^*=\arg\max_{r\in R}\mathrm{Sim}(o_i,r).
\end{equation}
If no raw label shares the same category as $o_i$, then $o_i$ is discarded from the computation.

\textbf{Step 3}: Normalize similarity within each category.
Because cosine similarities are not directly comparable across categories, we normalize each matched similarity using the global range of pairwise similarities within the corresponding category.
For category $c$, let
\begin{equation}
\mathrm{Sim}_{\min}^{(c)}=\min_{u,v\in \mathcal{T}^{(c)}} \mathrm{Sim}(u,v),
\mathrm{Sim}_{\max}^{(c)}=\max_{u,v\in \mathcal{T}^{(c)}} \mathrm{Sim}(u,v),
\end{equation}
where $\mathcal{T}^{(c)}$ denotes the set of all taxonomy labels in category $c$. Since self-pairs are included to represent the exact-match, $\mathrm{Sim}_{\max}^{(c)}=1$. We then define the exposure score of $o_i$ as
\begin{equation}
s_i
=
2\cdot
\frac{
\mathrm{Sim}(o_i,r_i^*)-\mathrm{Sim}_{\min}^{(c_i)}
}{
\mathrm{Sim}_{\max}^{(c_i)}-\mathrm{Sim}_{\min}^{(c_i)}
}
-1,
\qquad s_i\in[-1,1].
\end{equation}
This normalization evaluates exposure relative to the semantic range of the entire category rather than relative to a single reference label. The midpoint of the category range is mapped to $0$; thus, $s_i>0$ indicates residual exposure, while $s_i<0$ indicates that the obfuscated label is sufficiently dissimilar from the original one to be treated as successful obfuscation.

\textbf{Step 4}: Aggregate obfuscated labels at the original label level.
For each raw label $r_j$, let
\begin{equation}
A_j=\{\, o_i\in O \mid r_i^*=r_j \,\}
\end{equation}
be the set of obfuscated labels assigned to it. We define the per-label exposure as
\begin{equation}
e_j=
\begin{cases}
\max\!\left(0,\; \frac{1}{|A_j|}\sum_{o_i\in A_j} s_i\right), & \text{if } |A_j|>0,\\[8pt]
0, & \text{if } |A_j|=0.
\end{cases}
\end{equation}
This step first averages all obfuscated labels associated with the same original label, allowing distant or irrelevant obfuscated labels to reduce exposure by increasing ambiguity. The result is then clamped at $0$, so that the successful hiding of one original label cannot produce negative exposure that offsets leakage from another original label.

\textbf{Step 5}: Compute profile-level exposure and privacy gain.
Finally, we average over all original labels to obtain the overall exposure score:
\begin{equation}
\mathrm{Exposure}(R,O)=\frac{1}{N_R}\sum_{j=1}^{N_R} e_j.
\end{equation}
We then define the privacy gain as
\begin{equation}
\mathrm{PrivacyGain}(R,O)=1-\mathrm{Exposure}(R,O).
\end{equation}
A higher privacy gain indicates that the obfuscated profile preserves less information about the original profile.

\section{Decoy Combination Template}
\label{app:noise-template}

The \textsc{Decoy} action retains each selected sensitive unit and injects one  decoy value drawn from irrelevant data types within the same category (Section~\ref{sec:obfucation_actions}). Rather than blending the real and decoy values into a single sentence, \acronym{} rewrites the query into a slotted form and lists the fill-ins as separate query options, instructing the online AI Agent to answer each option independently under a labeled heading. These headings are what
let the response filter recover the answer to the real query and discard the decoy answers with a lightweight, hard-code parser, without a second model call. Prompt~\ref{lst:noise-template} shows the template instantiated on the running example of Figure~\ref{fig:pipeline0}, where the single noise unit ``therapist'' is paired with the decoy ``dentist'' ($N{=}1$, $V{=}2$).

\begin{lstlisting}[style=prompt, caption={\textbf{Decoy Template.} This is our Decoy template, instantiated on the running example
(Figure~\ref{fig:pipeline0}). The decoy unit ``therapist'' becomes \texttt{[slot
1]}; option~A carries the original unit and option~B the decoy unit. The chatbot returns
two labeled sections, and the response filter retains only the section for
option~A.}, label={lst:noise-template}]
The original query below has N=1 slots: [slot 1]. Each query option below (A, B) gives different values to fill in those slots. For EACH query option, form the complete query by filling in the slots with the given query option slot values. Then, give a full, independent answer to each filled-in query. Format your response with exactly 2 labeled sections:
"Answer for Query Option A: <your answer for A>",
"Answer for Query Option B: <your answer for B>".
Your responses must answer each query fully. Do not just list slot fill-ins, answer each query option.

The original query:
I am [REDACTED], and I was recently diagnosed with [Health]. I live in Palo Alto. Can you recommend a good [slot 1] near me?
Query Option A) [slot 1]=therapist
Query Option B) [slot 1]=dentist
\end{lstlisting}

\section{Adapting OPRO to Rule Optimization}
\label{app:opro}
OPRO's optimizer learns only from (solution, score) pairs, which works when the score is simple to interpret and cheap to compute~\cite{yang2024large}. Neither holds in our setting. Our score is composite: the objective in Eq.~\eqref{eq:opt-target} averages over queries and, within each query, over several coupled per-unit decisions, so a lower score cannot tell the optimizer \emph{which} decision went wrong. It is also expensive: each evaluation runs the Obfuscation Action Decider, the provider, and the scoring pipeline over the validation set.

\textbf{Per-query oracle.} Our first adaptation is to give the optimizer feedback at the level of individual decisions. This is possible because the action space is finite per query. For every training query, we score a subset of its action combinations before optimization begins, rather than all of them, since their number grows as $4^{M}$ with the number of sensitive units $M$. We add the score of each new combination the Decider proposes during optimization, and take the best combination recorded so far as the oracle $a^{\ast}(x)$. The oracle is therefore a lower bound on the true per-query optimum that tightens as optimization proceeds, which is why the ratio in Eq.~\eqref{eq:rule-quality} is capped at $1$. Queries whose oracle score is below $0.7$, for which no good combination is known, are excluded from $\mathcal{C}_r$. We report $q(r)$ both for the current mini-batch and cumulatively across rounds, and prompt the Rule Optimizer to weigh the cumulative statistics more heavily.

\textbf{Structured rule set.} Decision-level feedback must be attributed to something the optimizer can change, which fixes our second adaptation: $\mathcal{R}$ is a numbered list of rules rather than a free-form instruction. During optimization only, the Decider is prompted to report the rule it relied on for each decision, so feedback aggregates per rule, and the Rule Optimizer can keep, revise, remove, or add individual rules. The structure has a further benefit: an edit perturbs the policy only on the queries that touch the edited rules, which keeps successive proposals stable enough that one candidate per round suffices. At inference the reporting is switched off and the deployed Decider simply follows the rules.

\section{Outcome Region Summary of Single Actions}
\label{app:region-fractions}
Table~\ref{tab:region-fractions} summarizes numerically the per-query distributions of Figure~\ref{fig:per-query-kde} at maximum obfuscation strength ($\tau = 1.0$). For each obfuscation action, it reports the mean privacy and utility over all queries and the fraction of queries that fall into each of the four outcome regions defined in Section~\ref{sec:single-action}. The no-obfuscation baseline ($\tau = 0.0$) is included for reference.

\begin{table}[h]
    \centering
    \small
    \setlength{\tabcolsep}{4pt}
            \caption{\textbf{Obfuscation Action Outcome Region Summary.} Reported at maximum strength ($\tau = 1.0$). Region fractions partition queries at $\text{privacy} = 0.5$ and $\text{utility} = 0.5$ into \emph{success}, \emph{utility-collapse}, \emph{privacy-miss}, and \emph{failure}. Baseline ($\tau = 0.0$) is identical across actions.}
        \label{tab:region-fractions}
    \renewcommand{\arraystretch}{1.05}
    \resizebox{\columnwidth}{!}{%
    \begin{tabular}{l cc cccc}
        \toprule
        & \multicolumn{2}{c}{Mean} & \multicolumn{4}{c}{Region (\%)} \\
        \cmidrule(lr){2-3}\cmidrule(lr){4-7}
        Configuration & Priv. & Util. & Success & U-coll. & P-miss & Fail \\
        \midrule
        Baseline          & 0.21 & 0.90 & 21.1 & 0.4 & 78.0 & 0.5 \\
        \midrule
        \textsc{Redact}  & 0.59 & 0.63 & 39.4 & 25.2 & 31.2 & 4.1 \\
        \textsc{Abstract} & 0.70 & 0.65 & 51.2 & 25.4 & 20.1 & 3.3 \\
        \textsc{Replace}  & 0.62 & 0.67 & 41.9 & 24.2 & 29.6 & 4.3 \\
        \textsc{Decoy}   & \textbf{0.77} & \textbf{0.76} & \textbf{73.7} & \textbf{7.4} & \textbf{17.9} & \textbf{1.0} \\
        \bottomrule
    \end{tabular}%
    }
\end{table}

\section{Decision Rule Set}

\subsection{Oracle Seed Rule Set}
\label{app:oracle-seed}
The \textsc{Oracle} seed used to initialize rule-optimization
(Section~\ref{sec:seed}) is a set of the per-query oracle possibilities derived from Table~\ref{tab:oracle-selection}. Each importance bin inherits the optimal-action frequencies of the corresponding threshold row. Because that table credits every tied action, its rows sum to more than $100\%$. We therefore \emph{renormalize} each row into a probability distribution over the four actions, giving the priors below.

\begingroup
\setlength{\emergencystretch}{2em}%
\sloppy
\begin{enumerate}[label=\textbf{R\arabic*.}, leftmargin=*, itemsep=2pt]
    \item If the importance score of a sensitive unit is $> 0.8$, then
    $P = \{\textsc{Decoy}: 0.545,\ \textsc{Abstract}: 0.196,\ \textsc{Replace}: 0.156,\ \textsc{Redact}: 0.103\}$.
    The distribution is sharply peaked: \textsc{Decoy} is the default with high
    confidence.
    \item If the importance score is $> 0.6$ and $\le 0.8$, then
    $P = \{\textsc{Decoy}: 0.328,\ \textsc{Abstract}: 0.239,\ \textsc{Replace}: 0.233,\ \textsc{Redact}: 0.200\}$.
    A plurality favors \textsc{Decoy}, but the three alternatives bunch tightly
    ($0.20$--$0.24$), so the choice is moderately query-dependent.
    \item If the importance score is $\le 0.6$, then
    $P = \{\textsc{Decoy}: 0.277,\ \textsc{Abstract}: 0.249,\ \textsc{Replace}: 0.242,\ \textsc{Redact}: 0.232\}$.
    The prior is nearly uniform: no strong default, and the choice is highly
    query-dependent.
\end{enumerate}
\endgroup

\subsection{How the Rules Shape Action Selection}
\label{sec:rule-analysis}
We examine how the deployed decision rule set (Appendix~\ref{app:rules}) distributes the four obfuscation actions over sensitive units, and which rules account for each action. Applying the decision rule set to the $2{,}182$-query test set at $\tau=1.0$ yields $7{,}911$ unit-level decisions, distributed as \textsc{Decoy} $84.8\%$, \textsc{Abstract} $9.6\%$, \textsc{Redact} $5.4\%$, and \textsc{Replace} $0.2\%$. In the learned decision rule set, \textsc{Decoy} is the default choice: any sensitive unit that is the on-topic target of a query (the main topic, tool, product, place, person, work, or core concept the answer depends on) is assigned \textsc{Decoy} (\ruleref{20}), and any unit matching no other rule falls through to \textsc{Decoy} as well (\ruleref{21}). Because most detected units are integral to their query ($67.0\%$ have importance above $0.6$ and $31.5\%$ above $0.8$; see Appendix~\ref{sec:importance-validation-statistic}), \textsc{Decoy} handles the large majority of cases, chosen for $84.8\%$ of all units and for $91.8\%$ of importance $> 0.8$. The remaining three actions are supplementary, each reserved for a narrower class of units that \textsc{Decoy} serves poorly. \textsc{Redact} removes units that are sensitive but carry little task value, so deletion costs nothing: verbatim secrets, credentials, endpoints, and identifiers pasted in code, exact names copied from source text, and bare boilerplate phrases (\ruleref{1}, \ruleref{3}, \ruleref{4}, \ruleref{8}). Consistent with this, \textsc{Redact} is applied to $14.8\%$ of low-importance units (importance $< 0.3$) but only $4.3\%$ of high-importance ones (importance $> 0.8$). \textsc{Abstract} generalizes units that \emph{frame} a request rather than constitute it, where a coarser term still answers the query, such as a broad host language or platform rather than the specific API at issue (\ruleref{5}), or a generic role (\ruleref{11}), persona (\ruleref{13}), or background descriptor that wraps the real task (\ruleref{19}). Such units are predominantly of low importance. \textsc{Abstract} is applied to $20.4\%$ of all units below importance $0.6$, but only $3.7\%$ of those above $0.8$. Among the units to which \textsc{Abstract} is applied, the majority fall in \emph{Technology \& Computing} ($26\%$), \emph{Hobbies \& Interests} ($20\%$), and \emph{Education \& Occupation} ($16\%$). \textsc{Replace} substitutes a plausible same-type stand-in where the unit's surface form must be preserved but its exact value is not needed, such as an exact product model, year, or version (\ruleref{7}), or literature inspired by other original work (\ruleref{10}). During testing, these conditions match rarely: only $14$ units ($0.2\%$) receive \textsc{Replace}, confirming it as a specialized fallback rather than a workhorse.

\subsection{Deployed Decision Rule Set}
\label{app:rules}
For reference, we list the complete decision rule set deployed as \acronym{}, the
\textsc{Oracle}-seeded set learned under the balanced objective
(Section~\ref{sec:rule-analysis}). Rules are evaluated top to bottom; the first
matching rule fires, and rule~R21 is the default fallback. Action names are
capitalized. The marker $\hookleftarrow$ after each rule links back to the point
in Section~\ref{sec:rule-analysis} where that rule is discussed; conversely, a
rule number cited in that section (e.g.\ ``R5'') links forward to its entry here.

\begingroup
\setlength{\emergencystretch}{2em}%
\sloppy
\begin{enumerate}[label=\textbf{R\arabic*.}, leftmargin=*, itemsep=2pt]
    \item\label{rule:R1}\hypertarget{rule:R1}{} If the query rewrites, summarizes, translates, proofreads, or edits user-supplied text, and the sensitive unit is an exact copied person/company/organization/work title or a low-importance copied audience/participant noun from that text, assign \textsc{Redact}.~\ruleback{1}
    \item\label{rule:R2}\hypertarget{rule:R2}{} If the query is that same kind of source-text task, and the sensitive unit is a copied citation/reference, source/department/program label, background phrase, or other peripheral copied label, assign \textsc{Abstract}.~\ruleback{2}
    \item\label{rule:R3}\hypertarget{rule:R3}{} If the query contains a long pasted code block, log dump, or traceback, and the sensitive unit is a literal endpoint, table, schema, record-set name, log tag, env/account identifier, or actual secret/token value, assign \textsc{Redact}.~\ruleback{3}
    \item\label{rule:R4}\hypertarget{rule:R4}{} If the query contains a long pasted code block or traceback, and the sensitive unit is an exact copied import/include token or a local function/method/class/parameter name that is not the main external library/tool/error target, assign \textsc{Redact}.~\ruleback{4}
    \item\label{rule:R5}\hypertarget{rule:R5}{} If the query is coding/debugging and the sensitive unit is a broad programming language or platform name used only as the host medium, while a more specific construct/module/API/framework/error is the real target, assign \textsc{Abstract}.~\ruleback{5}
    \item\label{rule:R6}\hypertarget{rule:R6}{} If the query is review, listicle, marketing, or product-script writing and both a broad product family and a more exact year/model/version/trim are present, assign \textsc{Abstract} to the broad family label.~\ruleback{6}
    \item\label{rule:R7}\hypertarget{rule:R7}{} If the query is that same kind of product-writing task and the sensitive unit is the exact year/model/version/trim unit, assign \textsc{Replace}.~\ruleback{7}
    \item\label{rule:R8}\hypertarget{rule:R8}{} If the sensitive unit is only a bare production/deliverable phrase such as ``make a video,'' ``title,'' ``tags,'' ``script,'' ``description,'' or ``introduction,'' and it does not itself carry the substantive topic, assign \textsc{Redact}.~\ruleback{8}
    \item\label{rule:R9}\hypertarget{rule:R9}{} If the query asks for fanfiction, crossover, AU, parody, or content set in the universe/style of an existing work, and the sensitive unit is a source-work/franchise/universe reference used only as setting or aesthetic inspiration rather than the main subject, assign \textsc{Abstract}.~\ruleback{9}
    \item\label{rule:R10}\hypertarget{rule:R10}{} If the sensitive unit itself explicitly bundles a source work with fanfiction/crossover/AU/inspired-by wording, assign \textsc{Replace}.~\ruleback{10}
    \item\label{rule:R11}\hypertarget{rule:R11}{} If the query asks the assistant to adopt a broad role, profession, or generic persona only as a wrapper for another task, assign \textsc{Abstract} to that role/profession/persona label.~\ruleback{11}
    \item\label{rule:R12}\hypertarget{rule:R12}{} If the query is a first-person confession, persona/profile prompt, or family/intimate case setup, and the sensitive unit is a direct kinship term, a paternity/custody/test term, or a household-control/location phrase central to the setup, assign \textsc{Redact}.~\ruleback{12}
    \item\label{rule:R13}\hypertarget{rule:R13}{} If the query is that same kind of first-person family/intimate setup, and the sensitive unit is an adoption/status/background phrase or another low-importance social/background label with importance\_score <= 0.4, assign \textsc{Abstract}.~\ruleback{13}
    \item\label{rule:R14}\hypertarget{rule:R14}{} If the query is third-person fictional writing and the sensitive unit is only a broad kinship descriptor without an exact name or age, assign \textsc{Abstract}.~\ruleback{14}
    \item\label{rule:R15}\hypertarget{rule:R15}{} If the query is explicit sexual or nudity-focused creative writing, and the sensitive unit is an exact demographic/persona seed defining the subject, with importance\_score > 0.5, assign \textsc{Redact}.~\ruleback{15}
    \item\label{rule:R16}\hypertarget{rule:R16}{} If the query is hostile, threatening, or attack-slogan style, and the sensitive unit is the named target group, celebrity, character, franchise, or organization, assign \textsc{Abstract}.~\ruleback{16}
    \item\label{rule:R17}\hypertarget{rule:R17}{} If the query asks about evidence, history, worship, or beliefs regarding a specific named religious figure or deity, assign \textsc{Redact}.~\ruleback{17}
    \item\label{rule:R18}\hypertarget{rule:R18}{} If the sensitive unit is a long genetics/clinical marker string or list of alphanumeric haplogroup/assay identifiers, assign \textsc{Abstract}.~\ruleback{18}
    \item\label{rule:R19}\hypertarget{rule:R19}{} If the sensitive unit is only an output-control, answer-class, formatting, or style token and importance\_score <= 0.3, assign \textsc{Abstract}.~\ruleback{19}
    \item\label{rule:R20}\hypertarget{rule:R20}{} If the sensitive unit is the main named topic, tool, product, place, person, work, character, framework, or core concept needed to answer the query on-topic, assign \textsc{Decoy}.~\ruleback{20}
    \item\label{rule:R21}\hypertarget{rule:R21}{} Otherwise, assign \textsc{Decoy}.~\ruleback{21}
\end{enumerate}
\endgroup

\section{User-Level Evaluation Details}
\label{app:multiturn}

\subsection{User Sampling}
\label{app:multiturn-buckets}
We treat all queries that share a hashed IP address in WildChat~\cite{zhao2024wildchat} as coming from one user. We bucket English-language users by their total number of queries across all of their conversations (1--5, 6--10, 11--15, 16--20, 21--25) and sample 20 users at random from each bucket, yielding 100 users, 480 conversations, and 1,297 queries. 91.7\% of English WildChat users send at most 19 queries, so the buckets cover the bulk of the population. The provider's conversation history resets at each new conversation.

\subsection{Dependency Labeling}
\label{app:multiturn-labeling}

We label each non-first query with the ``New Ask or Follow up'' annotator prompt from Appendix~A.20 of Huang \etal{}~\cite{huang2026llms}, giving GPT-5~\cite{openai_gpt5} the full preceding conversation as context. The prompt assigns \emph{new\_ask}, \emph{feedback} (the user reacts to or requests changes to an earlier response), or \emph{no\_feedback} (the user continues the same task without evaluating it). We treat \emph{new\_ask} as independent and the other two as dependent, giving 416 independent and 401 dependent queries (163 feedback, 238 no\_feedback).

\end{document}